\documentclass[sigconf]{acmart}

\AtBeginDocument{%
  }

\copyrightyear{2025}
\acmYear{2025}
\setcopyright{acmlicensed}\acmConference[CHI '25]{CHI Conference on Human Factors in Computing Systems}{April 26-May 1, 2025}{Yokohama, Japan}
\acmBooktitle{CHI Conference on Human Factors in Computing Systems (CHI '25), April 26-May 1, 2025, Yokohama, Japan}
\acmDOI{10.1145/3706598.3714182}
\acmISBN{979-8-4007-1394-1/25/04}

\usepackage{xspace}
\usepackage{multirow}
\usepackage{colortbl}

\usepackage{natbib}

\definecolor{sigcell}{RGB}{253, 253, 150}

\begin{document}

\title{Virtual Voyages: Evaluating the Role of Real-Time and Narrated Virtual Tours in Shaping User Experience and Memories}
%\title{I'm Off for a While: Exploring the Effect of Being Live and Storytelling on Presence, Place Attachment, and Affect within Virtual Tourism}
%Virtual Voyages: Exploring the Effects of Narrative and Live Experiences on Virtual Tourism
%%Measuring Experience Feel/User Experience 

\author{Lillian Maria Eagan}
\email{eagli237@student.otago.ac.nz}
\orcid{0009-0005-4609-0173}
\affiliation{
    \institution{University of Otago}
    \city{Dunedin}
    \country{New Zealand}
}

\author{Jacob Young}
\email{jacob.young@otago.ac.nz}
\orcid{0000-0002-1180-4114}
\affiliation{
    \institution{University of Otago}
    \city{Dunedin}
    \country{New Zealand}
}

\author{Jesse Bering}
\email{jesse.bering@otago.ac.nz}
\orcid{0000-0002-1501-702X}
\affiliation{
    \institution{University of Otago}
    \city{Dunedin}
    \country{New Zealand}
}

\author{Tobias Langlotz}
\email{tobias.langlotz@cs.au.dk}
\orcid{https://orcid.org/0000-0003-1275-2026}
\affiliation{
    \institution{Aarhus University}
    \city{Aarhus}
    \country{Denmark}
}
\additionalaffiliation{
    \institution{University of Otago}
    \city{Dunedin}
    \country{New Zealand}
}

\renewcommand{\shortauthors}{Eagan et al.}

\begin{abstract}

Immersive technologies are capable of transporting people to distant or inaccessible environments that they might not otherwise visit. Practitioners and researchers alike are discovering new ways to replicate and enhance existing tourism experiences using virtual reality, yet few controlled experiments have studied how users perceive virtual tours of real-world locations. In this paper we present an initial exploration of a new system for virtual tourism, measuring the effects of real-time experiences and storytelling on presence, place attachment, and user memories of the destination. Our results suggest that narrative plays an important role in inducing presence within and attachment to the destination, while livestreaming can further increase place attachment while providing flexible, tailored experiences. We discuss the design and evaluation of our system, including feedback from our tourism partners, and provide insights into current limitations and further opportunities for virtual tourism.

%Immersive technologies are increasingly being considered for their potential to replicate or enhance existing tourism experiences, especially to share inaccessible or distant environments with visitors who otherwise would not be able to experience them. While the academic literature suggests substantial interest in this area,  there have been few, if any, controlled experiments to study how users perceive virtual tours of real-world locations. In this paper we present an initial exploration of virtual real-world tourism focusing on the effects of real-time experiences and storytelling on presence, place attachment, and user memories of the real-world destination. Our results suggest that prerecorded and live experiences could both be valid approaches, and that storytelling can enrich the experience for users in several ways. We discuss the design and evaluation of our system, including feedback from domain experts, and provide insights into current limitations and further opportunities for virtual tourism. 

\end{abstract}

\begin{CCSXML}
<ccs2012>
   <concept>
       <concept_id>10003120.10003121.10011748</concept_id>
       <concept_desc>Human-centered computing~Empirical studies in HCI</concept_desc>
       <concept_significance>500</concept_significance>
       </concept>
   <concept>
       <concept_id>10003120.10003121.10003124.10010866</concept_id>
       <concept_desc>Human-centered computing~Virtual reality</concept_desc>
       <concept_significance>500</concept_significance>
       </concept>
   <concept>
       <concept_id>10003120.10003121.10003124.10010392</concept_id>
       <concept_desc>Human-centered computing~Mixed / augmented reality</concept_desc>
       <concept_significance>500</concept_significance>
       </concept>
   <concept>
       <concept_id>10003120.10003121.10003122.10011749</concept_id>
       <concept_desc>Human-centered computing~Laboratory experiments</concept_desc>
       <concept_significance>500</concept_significance>
       </concept>
 </ccs2012>
\end{CCSXML}

\ccsdesc[500]{Human-centered computing~Empirical studies in HCI}
\ccsdesc[500]{Human-centered computing~Virtual reality}
\ccsdesc[500]{Human-centered computing~Mixed / augmented reality}
\ccsdesc[500]{Human-centered computing~Laboratory experiments}

\keywords{Virtual Tourism, Virtual Reality, Telepresence, Mixed Reality, Presence, Storytelling, Narrative, User Study, Place attachment, Empirical Study, 360 Panorama, Spatial Audio}

\begin{teaserfigure}
\centering
  \includegraphics[width=0.995\textwidth]{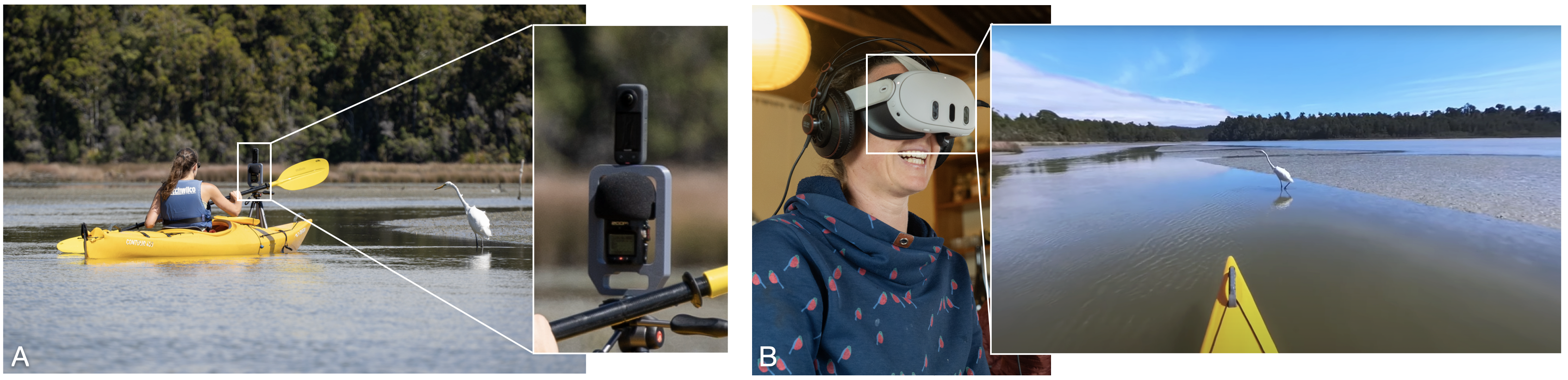}
  \caption{We present Virtual Voyages, an exploration of user experiences in virtual tourism. In partnership with local tourism operators, we created a virtual kayaking experience that takes users on a virtual tour of Ōkārito Lagoon in New Zealand. (A): This experience was captured using a 360$^\circ$ camera and ambisonic microphone mounted to the front of a kayak. (B): Users can view this experience in a VR head-mounted display without narrative, with a prerecorded narrative, or talking live to a tour guide.}
  \Description{Three images arranged horizontally in a line. The leftmost image shows a woman in a kayak in the middle of a lagoon with a large white bird next to her. A white box outlines a 360 degree camera and spatial microphone mounted to the front of the kayak; white lines come from this box and frame the centre image to show that it is a close-up image of this camera setup. The rightmost image shows an image from the perspective of the camera; the front of the kayak can be seen coming out from the bottom edge, with the large white bird in the centre of the image.}
  \label{fig:teaser}
\end{teaserfigure}

\received{12 September 2024}
\received[revised]{12 March 2009}
\received[accepted]{5 June 2009}

\maketitle

\section{Introduction}
Virtual Tourism (VT) envisions a new way to travel to familiar or unfamiliar places without leaving home. Compared to traditional tourism based on physical travel, VT facilitates experiences in a more carbon-friendly way, offers new experiences to those unable to travel, and reduces the strain on communities feeling the effects of over-tourism. Concerns that VT will create competition for traditional travel~\cite{Voronkova2018, Cheong1995} overlook its potential to support, complement, or enhance in-person tourism experiences~\cite{Beck2019, Verma2022, Lu2022}.

Tourism and hospitality literature has already started to explore immersive technology, with specific uses cases for travel destination marketing, tours of historic sites~\cite{liestolAppianWayStorytelling2014}, and how to enhance museums with virtual content~\cite{siang2019}. However, the use of digital 3D reconstructions~\cite{Bastanlar2008, Jego2019, Kyrlitsias2020} prevent dynamic features such as weather, people, and wildlife~\cite{stotko2019slamcast, Young2020, Mohr2020}. With recent advancements in the quality, portability, and accessibility of panoramic cameras, several livestreaming solutions have been proposed for tourism applications ~\cite{alegro2023introducing}, though these have limited interactivity between tourist and guide. Furthermore, much of the literature focuses on the technical difficulties of virtual tourism, neglecting investigations into how to design a good \textit{experience} rather than a good piece of software~\cite{Heeter2003, Banos2004, Katkuri2019}.

In this work we explore the concept and characteristics of Virtual Tourism, specifically the role that livestreams and narrative may play on the quality of the experience, place attachment to the destination, and the sense of presence within it. Through a co-design process with local tourism providers, we prototyped several Virtual Reality (VR) kayaking experiences that guide users through Ōkārito Lagoon in New Zealand. This system was used in a study to explore user perceptions of virtual tours, with our results suggesting that the inclusion of narrative content (e.g. story-driven material) can result in a richer, more fondly remembered virtual reality tourism experiences that increases the user's sense of spatial presence within and place identity with the destination. We also found that the perception of watching the experience live had no significant effect on the user's presence, but a live experience could increase their attachment to the destination. These findings are of increasing relevance not only to physical and digital tourism but also to practitioners in telepresence, immersive technologies, and human-computer interaction. Finally, we provide qualitative feedback from our participants and tourism partners to help guide the future development of virtual tourism systems.

Our main contributions are as follows:
\begin{itemize}
    \item A novel VR tour of Ōkārito Lagoon that integrates 360$^\circ$ video, ambisonic audio, and live narration;
    \item Our approach for designing virtual tourism experiences in collaboration with established tourism operators;
    \item Results of a user study indicating that narrative in a tourism experience can increase user's presence within and attachment to the destination through a higher-quality touristic experience, and that sharing the experience live may increase this place attachment further;
    \item Discussion and feedback from domain experts to guide the development of future virtual tourism experiences.
\end{itemize}

%\section{Related Work}

%Virtual tourism spans multiple disciplines in its own right - Human-Computer Interaction \cite{Wolf2023}, Virtual and Augmented Reality \cite{Verma2022, Afsarmanesh2000, Zhang2020}, Presence and Telepresence \cite{Hoang2023, Yazaki2023, Bafadhal2019, Manabe2020}, Environmental Psychology \cite{Cheng2022, Huang2023}, and Hospitality/Tourism \cite{Cheng2022, Lu2022, Yang2023}. Each of these fields offers methodology, concepts, and theories that can help advance our understanding of what technology-supported `new' tourism might look like. Researchers are already contemplating what value it can provide, and, similarly, what it cannot do. In this section, we present relevant concepts, prototypes and studies that have explored virtual tourism.

\section{Related Work}
Technology is revolutionising how we travel in exciting new ways \cite{Olsen2000, Xiang2015, Ukpabi2017, Torabi2022, Jalilvand2024}, from digital tourism~\cite{Benyon2014} to technology-supported tourism~\cite{Yoo2017} to completely virtual tourism~\cite{Lu2022, Verma2022}. These discordant terms - sometimes used interchangeably - pose a problem for conducting rigorous multidisciplinary research~\cite{Mura2016, Beck2019}.

%communicating ideas across the relevant disciplines~\cite{Mura2016, Beck2019}.

%Digital tourism is often broad, encompassing any applications that offer digital support of the tourist experience \cite{Benyon2014}. On the other hand, virtual tourism is refers to an increasingly spcific concept: the use of VR in tourism context.  

For example, virtual tourism can be used generally, similar to ``technology-supported tourism''~\cite{Yoo2017} or ``digital tourism''~\cite{Benyon2014}. For the scope of this work, we follow the notion that virtual tourism is a form of tourism that specifically uses Virtual Reality (VR) to virtually transport people to remote places. This may be accomplished using a wide range of technology, including hand-modeled worlds, 3D reconstructions, and immersive videos or livestreams.

%Yet technologies, measurements, psychological impacts, and ethics have never been combined in a study like this... 

%despite sharing the same goal: providing a memorable experience for virtual tourists.

%One of the most promising and underutilised technologies that could be used to create meaningful 360$^\circ$ real-captured virtual reality tourism experiences is livestreaming telepresence systems. 

\subsection{Enabling Technologies}
Early applications of VR tourism offered users a variety of virtual environments from cultural heritage sites to museums~\cite{Bastanlar2008, Book2003, Guttentag2010}, all of which relied heavily on static 3D reconstructions and handmade 3D models. While visually appealing, 3D reconstructions are computationally expensive, are prone to errors (particularly in outdoor environments)~\cite{Mohr2020}, and are not feasible in real-time for large environments~\cite{Young2020}.

More recently, real-world content captured using 360$^\circ$ video has become popular thanks to improving camera quality, affordability, and mainstream usage~\cite{Young2019, Tang2017}. This suggests that a new era of immersive videos is imminent, yet currently, content for VR in tourism, regardless of 3D, 2D or panoramic, is still mainly prerecorded, passive and lacks depth in terms of narration.

This gap of livestreaming has been identified and in particular since the pandemic livestreaming is increasingly relevant in tourism contexts~\cite{Deng2021, Lin2022, Shen2022}, promising a new travel phenomenon that is ``simultaneous, immediate, social, and vicarious''~\cite{Deng2019}. Howeverm,  while Youtube and Twitch livestreams demonstrate interest from worldwide markets, they lack a sense of immersion and can disrupt viewership with significant latency, limiting the interaction between the streamer and their audience. The field of telepresence goes beyond livestreaming to discover what other interactions are possible to make people feel like they're physically in a distant place~\cite{Deng2019}. Recent innovations in telepresence include 360$^\circ$ video)~\cite{Tang2017}, independent viewing~\cite{Young2019, Young2020}, 3D capture~\cite{Young2020, stotko2019slamcast}, and embodiment~\cite{Orts2016}, all of which have proven capable of increasing this sense of being somewhere else. 

However, telepresence has scarcely been applied to tourism despite this obvious link, with the few works that do exist focusing on the technologies involved rather than the quality of the user's experience~\cite{Manabe2020, Yazaki2023}. A recent literature review has argued that further research is necessary for academics and practitioners to understand the role of live streaming in tourism and hospitality~\cite{Lin2022}. Their work identifies three research priorities: diversified geographical or cultural contexts, collaboration with stakeholders, and theoretical and methodological advancement. This study aims to address all three of these priorities.

%The range of VR Tourism Experiences (VRTEs) is vast, from digital tourism~\cite{Benyon2014} to technology-supported tourism \cite{Yoo2017} to virtual tourism \cite{Lu2022, Verma2022}. Despite similar names, these concepts are often seen as distinct, and discrepancies of definition and the interchangeable use of terms poses a problem for communicating findings and ideas across the relevant disciplines \cite{Mura2016, Beck2019} despite sharing the same goal: providing a memorable experience for virtual tourists. 

%One possible approach for studying the user experience of virtual tourists may be utilising 360$^\circ$ real-world content for on-site virtual reality tourism experiences (VRTEs), which has become increasingly inexpensive and accessible to the general public \cite{Beck2019}. By leveraging immersive technologies we can analyse key experience characteristics that maximise presence, meaning, and memorability. The key question being: which elements are most critical to the experience of presence \cite{IJsselsteijn2003}, and how can these contribute to a meaningful touristic experience?

\subsection{Virtual Reality Tourism Experiences}
Central to VR is the concept of presence, or the sense of ``being there'' in the virtual environment~\cite{Biocca1997}. This concept naturally aligns with the defining feature of tourism - visiting far away places - making it a crucial element of meaningful VR tourism experiences. Consequently, much of the virtual tourism literature investigates how various experiential elements influence presence, revealing a complex web of factors and features to consider. Yet findings can vary depending on methodology, level of immersion, or application~\cite{Jung2016, Tussyadiah2018, Beck2018, Beck2019, Melo2022, Fan2022, Pantelidis2024, Song2024}. 
%This makes it difficult to actually contribute something to the literature in a generalizable way, only for what feels like niche applications. 

Studies on virtual tourism tend to agree there is a positive link between immersion and presence~\cite{Beck2019, Rainoldi2018}, with immersion referring to the attributes of the technology itself rather than the user's experience of it~\cite{Slater1997}. While some studies focus intently on maximising immersion, others have shown the immersion-presence relationship is not always one-to-one. This highlights a tendency for VR studies to overemphasise immersion at the expense of other crucial experience elements such as the media content~\cite{Banos2004}.

%Some works define virtual tourism broadly, encompassing tourism destination content viewable on smartphones, tablets, or desktops \cite{Verma2022, Lu2022}. This differs from other literature, which limits the scope of virtual tourism to applications of Virtual Reality (VR) in a tourism context \cite{Tussyadiah2018, Merkx2021} (VRT). In the following work, we refer to the latter definition, to contribute to the growing body of work that explores virtual reality systems and their capability to take tourists places virtually and create meaningful experiences. 

%\citet{Beck2019} contributed a classification of VRT according to level of immersion: Non-immersive (content displayed on a conventional desktop display), semi-immersive (content displayed on large monitors, walls, or floors), and fully-immersive (content viewed in a VR headset), with immersion referring to the attributes of the technology used rather than the experiential phenomenon of presence~\cite{Slater1997}.

Existing fully immersive VT studies primarily focus on the ``pre-travel'' phase of tourism, emphasising marketing and promotion of on-site experiences~\cite{Yung2021.2, Fan2022, Liu2023} rather than designing something that can stand on its own merit as a satisfying experience~\cite{Beck2019} or possibly even provide a viable substitute for physical travel~\cite{Guttentag2010}.

The existing bias in the literature may be due to the barriers of VR content creation or the desire to produce literature that serves organisational goals~\cite{Beck2019}. Understandably, investigations catered towards economic opportunities can help express the potential impact of this field to the entire tourism industry, however this fails to capture the potential from a user experience perspective. For example, pre-travel studies often measure purchase intention or intent to visit, which don't necessarily tell you how enjoyable, satisfying, or memorable an experience is~\cite{Fan2022, Yung2021.2}.% This is echoed by \citet{Pine1998} who commented on the trend for companies to simply wrap experiences around their existing products with the motivation to sell them better \cite{Pine1998}. This fails to properly explore the potential of experiences on their own, outside of inherent economic motivations.

Lastly, there is an increasing interest in documenting the socio-psycholoigcal effects of virtual tourism on visitors. Currently, scholars believe virtual tourism is unlikely to satisfy the socio- psychological needs that traditional tourism meets in regards to meeting people and forming bonds~\cite{Tjostheim2022}. However, as experiences improve in quality and offer opportunities for social engagement, a deeper understanding of the socio-psychological impacts will be necessary. 

\subsection{The Realms of Tourism} 
In their seminal work, \citet{Pine1998} propose that experiences, even ones not tied to physical goods or services, can be their own distinct economic offering. These experiences can be expressed in four broad categories which \citeauthor{Pine1998} call the ``Four Realms of an Experience'': esthetics, entertainment, education, and escapism.

This concept has revolutionised the fields of marketing, hospitality, and, especially tourism~\cite{Song2015}, where it is known as the Four Realms of the \textit{Tourism} Experience. \citet{Oh2007} applied this theory to the tourism sector as a way of assessing a destination's value and added four new tourism-specific factors: arousal, memory, quality, and satisfaction. However, research on how the Realms, including the additional factors, might apply to \textit{virtual} experiences remains scarce~\cite{Lee2020}. 

While this framework provides a foundation for thinking about and studying experiences, to focus on economic viability limits our ability to understand what other value an experience can provide; not to mention economic value is highly subjective and dependent on one's circumstances~\cite{Mitchell1999, Sesini2023}. Therefore, we consider metrics beyond economic value to measure the quality of memorable VRTEs~\cite{pizam2010creating}.

\subsection{Place Attachment and Memory}
Tourism literature has increasingly investigated the environmental psychology phenomenon known as \textit{place attachment}, which is defined as the physical, social, or emotional bond between person and place. Place attachment can reveal the nature of human-place relationships such as behavioural intentions and how meaning is ascribed to physical environments~\cite{Dwyer2019}. Studies have found that place attachment is facilitated by experiences that are memorable, satisfying, and that enhance a person's purpose and meaning in life~\cite{Vada2019}. However, it is unclear how place attachment might apply to a virtual destination rather than a physical one.

%A small but growing number of studies have begun to investigate place attachment in virtual reality, suggesting that place attachment is an important element of VRT but the mechanisms remain unclear \cite{Wang2022, Kim2023.2, Pantelidis2024, Pantelidis2018}. Specifically, no studies have investigated which experience attributes may facilitate place attachment in virtual reality/environments using a controlled experiment. 

%%CITE WANG 2022 While some literature has explored the effects of virtual reality tourism on place attachment, such as \cite{Wang2022}, which offered users a variety of prerecorded panoramic videos of the Yellow Crane Tower in Wuhan, China, our study is the first to integrate a narrative into a real-time tourism experience. 

Environmental psychology literature also posits \textit{memory} as a key phase of travel~\cite{Fridgen1984}, suggesting that post-travel satisfaction is a key part of a touristic experience. Studies have supported the link between positive experiences and positive memories in on-site and virtual tourism alike, suggesting that positive memories reflect that a positive experience was had~\cite{Kim2022}. The role of memory in tourism is significant, however its mechanisms are not yet well understood~\cite{MARSCHALL20122216}, especially in a virtual context.

Recent literature has focused on which aspects of an experience can positively affect a tourist's memory of it~\cite{Tung2011}, and suggest that a person's \textit{affect}, or their underlying \textit{pleasure} (positive emotion) and \textit{arousal} (psychological alertness)~\cite{russell1989affect}, can significantly distort their memory of the experience~\cite{Zurbriggen2021}, even in virtual environments~\cite{Mancuso2023}.

%%By replicating a touristic experience virtually we can present an idealised version of it, removing negative aspects such as poor weather and long travel times, potentially resulting in an experience that is more fondly remembered.

%Finally, one may be wondering how a VRTE may best be suited to facilitate positive affect and place attachment(, therefore creating meaningful experiences). Trends in the literature have repeatedly focused on three experience characteristics: narrative, presence, and immersion. The following sections briefly review these and propose an approach for a novel technology system for creating and studying meaningful VRTEs.  

%\subsection{Narrative}
%Both place attachment and memory offer ways to quantify/study value beyond economics. What aspects of experiences increase PA and positive memory?

Narrative has been suggested as one way to make these experiences more meaningful~\cite{Cao2021, Spielmann2018}, with significant links found between narrative and immersion, narrative and emotional response, and narrative and presence~\cite{Gorini2011}. However, this has been contradicted in other literature~\cite{Weech2020}, making narrative's role in virtual tourism unclear. Further study is required to determine if a VRTE's narrative plays an important role in the overall satisfaction of its users.

\subsection{Ethical Considerations}
Given the enormous potential for VR to change the tourism industry at-large, it is important to consider the ethical aspects of this work. 
In recent years the tourism industry has made a marked shift towards regenerative tourism, or sustainable tourism that aims to positively impact local communities and environments~\cite{Bellato2023}. This values-based refocusing signals that the entire industry is exploring innovations that align with these goals. VR and VRT present a less-carbon dependent form of travel, which could increase its popularity in coming decades~\cite{Talwar2023}. 
Yet, this raises the question of security and privacy considerations for the visitor and user experience~\cite{Meena2024}. While our work is not a specific investigation of the ethics of VRT, in the following we seek to document the prominent ethical considerations that pertain to our work. This includes the impact of meaningful VRTEs on visitors, physically and psychologically, and the consequences for tourism operators. 

\subsection{Summary of the Literature}
Overall, there is a vast number of works in the literature discussing the potential of VR for tourism. Many of those focus on marketing and less actually built and explored systems that are truly immersive. Most of those used offline reconstructions that lack dynamic details and narration (cp. experience beyond technology~\cite{IJsselsteijn2003} are often poorly reported. Instead, this work explores, in partnership with actual stakeholders, immersive telepresence techniques using live streaming and immersive VR technologies directly addressing the research challenges identified by Lin et al. ~\cite{Lin2022}. Our goal is to create a better understanding of the role of live streaming and narration which so far is not understood  (cp. lack of theoretical insights ~\cite{Lin2022}).

\section{Creating A Virtual Kayak Experience}
To design this research we collaborated with tourism providers who offer guided kayaking tours in Ōkārito, the largest unmodified wetland in New Zealand that has been identified as an important habitat and nesting site for many significant endemic birds. In order to adapt their in-person experience for virtual reality we participated in their guided tours, regularly visited the area, and performed extensive tests and research over an extended period. In the following we describe our process of working with our tourism partners to create the virtual experience that became the plaform for our research on narrative and real-time experiences. 
%This required: six multi-day on-site visits with our tourism partners, the development of a novel technology system for live guided tours, editing the experience using 360-video and ambisonic sound, and writing, rehearsing and recording an engaging narrative.  

\subsection{Early Engagement} 
%The core part of the project took place over a twelve-month period; however, early engagement and initial discussions date back 30 months. 
When first engaging with our partners the tourism sector was still facing the COVID-19 pandemic, with the subsequent under-tourism threatening the economy of their tourism-dependent region. However, it became clear that over-tourism is also problematic as it threatens the health of the region's the native forests, landscapes and wildlife. We imagined how virtual tourism solutions could offer a pathway to diversify the sector, perhaps decreasing dependence on physical travel while reducing stresses on the environment. 

We considered other tourism operators for this work, however kayaking tours offered a few specific benefits:  
%We explored multiple experiences for our virtual tourism content, including bird tours, guided hikes, and glacier helicopter rides. However, virtualising kayaking experiences emerged quickly as a promising first case study on virtual tourism for this region/project for several reasons:
\begin{itemize}
    \item Kayaking naturally restricts the user's movement to the water and the seat of the kayak, limiting the need for wide-area reconstructions and providing a natural path for users to move through.
    \item Tandem kayaks provide a believable means of moving the user's viewpoint through the environment, potentially minimising the effects of simulator sickness.
    \item The hull of a kayak provided a natural mounting and storage solution for technical equipment.
\end{itemize}  

%Additionally, we felt our research aims aligned with their business ethos: to provide their customers with a valuable, meaningful experience:
%\begin{quote}
%    \textit{``We deliberately limit the size of the business in order to achieve a really high quality of experience for people. We often suggest people don't go outside, even when we can send them out. [..] We don't want to send people out who won't enjoy it."}
%\end{quote}
As the lagoon's gatekeepers, the tourism operators were excited by the opportunity to deliver their product to a wider market of people, especially those unable to travel physically due to finances, health, visas, or environmental concerns. 

%\subsection{Understanding Current Practices and Context}
A key differentiator for our partners was how they catered their tours to each customer based on their preferences and the unpredictability of nature:
\begin{quote}
    \textit{``We like a really high degree of customisation, so that it doesn't become just generic. You need to know what kind of personality types you're dealing with, to know what kind of tour somebody wants to kind of cater it to their preferences. We've had times where I paddled parallel to a white heron for five minutes... But if somebody says they don't care about birds, you're not going to show them that. So it's not always the same experience.''}
\end{quote}
This further emphasised the specific suitability of guided kayak tours for our research and supports the idea that live tours, even in virtual formats, may provide a more curated and repeatable experience than pre-recorded tours.

\begin{figure*}
    \centering
    \includegraphics[width=\textwidth]{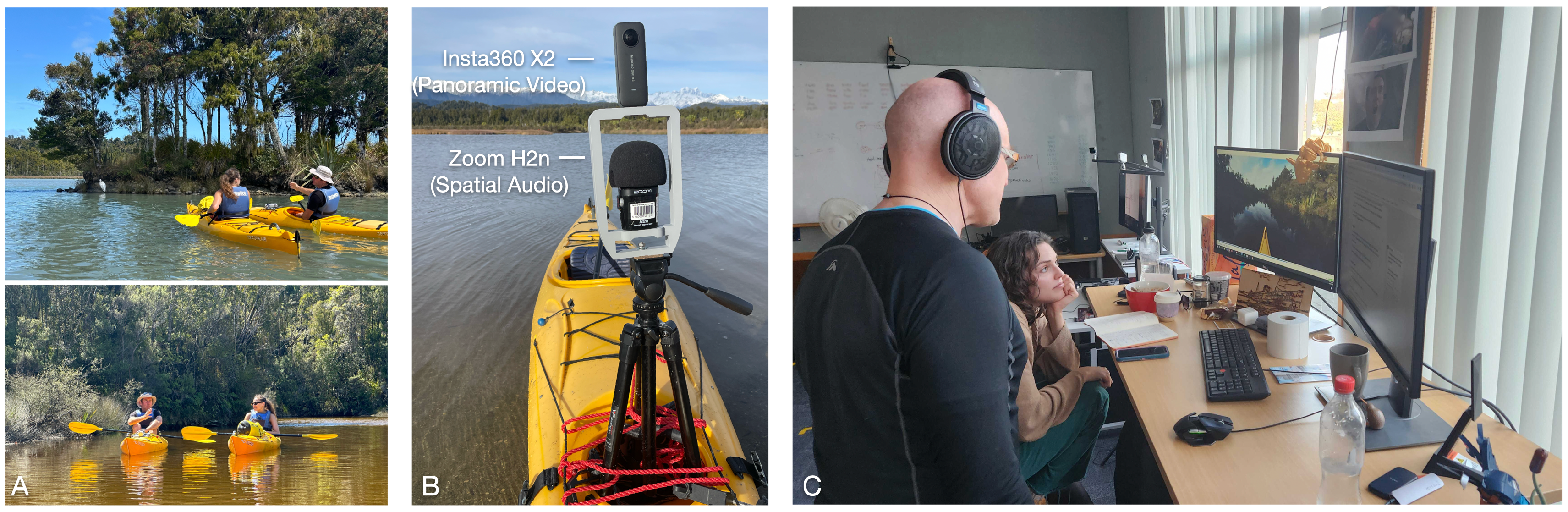}
    \caption{Our process for designing the virtual kayaking experience. (A) Early visits to the location to understand the environment and how a guide usually operates their kayak tours. (B) A 15-minute ride into one of the channels was recorded using an Insta360 One X2 360$^\circ$ camera and a Zoom H2n attached to the front of the kayak with a custom 3D-printed mount. (C) We worked with a paid actor and our tourism partners to write a script which was used for the narrative component of our user study.}
    \Description{Four images arranged in three columns, with the first two images stacked on each other. The first column shows a photo of two people in kayaks on a clear lagoon (top), and a man seated in a chair wearing a Meta Quest 3 to revisit the lagoon virtually. The second column shows a close-up view of the front of a kayak; a 360 degree camera and spatial microphone are attached to a tripod sticking out of a small hole in its hull. The third column shows two people looking at a computer monitor that is displaying an image of a kayak on the lagoon.}
    \label{fig:Setup}
\end{figure*}

%I THINK THIS IS COVERED IN EARLY ENGAGEMENT ABOVE: We also observed that it is not uncommon for the guides to share a double kayak with their customers, particularly those who would be unable to paddle their own. This effectively reduces the tourist's autonomy as only the guide has access to the steering mechanism, making it feasible to design experiences around pre-determined paths.

%With each visit, we scouted locations and tested perspectives for the camera mount on the kayak. 

\subsection{Final Experience Design: Feedback and Story Design}
Through our testing we concluded the front compartment of the kayak created the most realistic perspective for the user, with the user in the front and the guide in the back, as seen in \autoref{fig:Setup}. This also informed our decision to use a 360$^\circ$ camera as we do not need to support unconstrained movement, and it has the added benefit of capturing dynamic elements such as water and wildlife which aren't currently supported by complex 3D reconstructions~\cite{Mohr2020}.

Finally, we determined an essential characteristic of the experience was the environmental soundscape. Thus, we altered our custom camera rig to accommodate a dedicated recorder for spatial audio, as seen in \autoref{fig:Setup}B). 

The experience was recorded over two days. Several select recordings were integrated into an early prototype of our VRTE, which we showed to our tourism partners within a head-mounted display (HMD).% Though these early prototypes lacked an integrated narrative voice-over, our partners were able to offer feedback based on the visuals and soundscape.% It was the first time our partners had experienced VR and so early opinions may have been biased by a novelty effect; yet, they were completely immersed, not even responding to questions while viewing the content in the HMD. 

These informal evaluations revealed several key findings. First, the resolution of both the camera and display made specific bird identification - a key facet of our partner's normal guiding duties - difficult. They also indicated that on the open lagoon, the visuals struggled to convey the vastness of the space, which probably can be attributed to the reduced field of view of current HMDs. Third, certain visual elements of the scene seemed inconsistent with normal, real-life vision. For example, in the smaller channels the mirror-like surface of the water reflected objects like clouds and trees unnaturally, making it difficult to identify the waterline. Finally, the sound of the kayak itself was sometimes too loud in comparison to the ambient and atmospheric sounds from the environment. Despite these shortcomings, our partners repeatedly stated how impressive the visual and audio quality was, mentioning that they felt present in the channels of the lagoon and disconnected from their physical environment. They saw immediate value in giving potential tourists a sense of what to expect before an actual visit. 

\begin{figure*}
    \centering
    \includegraphics[width=\textwidth]{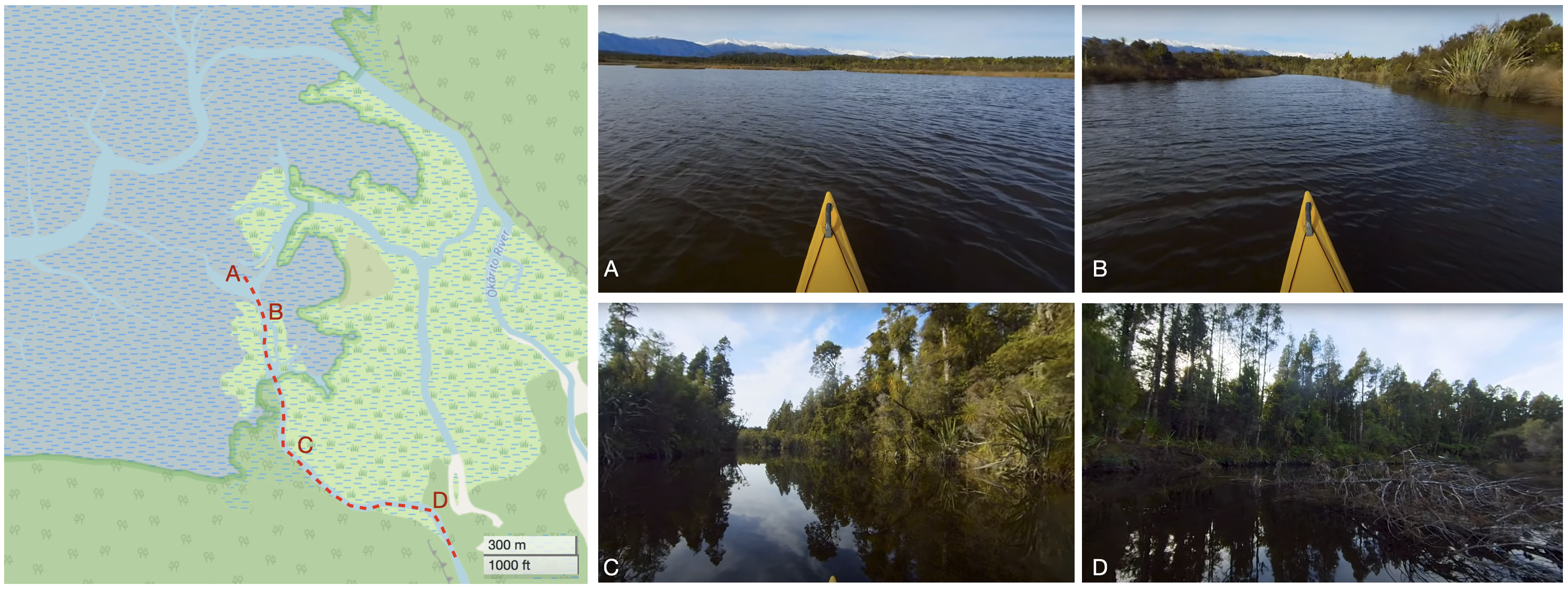}
    \Description{Five images arranged in a grid. The leftmost image, almost the size of the other four, shows a map largely dominated by water. There is a line heading towards the bottom-right corner with labels A, B, C, and D placed along its length. The four images on the right show first-person perspectives from a kayak on a lagoon; each has a letter in the corner showing which point on the map it was taken from. The first image shows a wide-open view of a lagoon, with each subsequent image becoming more closed-in until the fourth image shows the kayak in a narrow channel with native flora on both banks.}
    \caption{(Left): The route followed during the virtual tour shown to participants. (Right): Screenshots taken from this experience at points A, B, C, and D along this route.}
    \label{fig:StudyPath}
\end{figure*}

We were then able to finalise the content for the study. One 15-minute, single-take recording was selected and a script was written to match the video's content; the route taken during the recording is shown in \autoref{fig:StudyPath}. In an effort to increase external validity, the narrative for the 15-minute VR experience was adapted from the provider's existing in-person guided tour which focuses on the area's ecological, cultural and historical significance. Our filming location was thus selected for its varied scenery, abundant bird life, and wind protection.

\section{A Study on Live Virtual Tourism Experiences}
%Based on the feedback and suggestions from our tourism partners we developed the final study design and prototypes. 

The main purpose of the study was threefold:
\begin{enumerate}
    \item To understand the effects and perception of a live, real-time, virtual tourism experience
    \item To assess how narrative-driven virtual tourism experiences contribute to touristic experiences for users
    \item To gather general feedback on our idea and implementation of virtual tourism from novice users and professionals from the tourism sector. 
\end{enumerate}

\subsection{Study Design}
To achieve our goal of exploring live virtual tourism experiences, we conducted a between-subjects experimental study, collecting feedback on different implementations of a 15 minute virtual tour of an ecologically significant wetland. This was captured using $360^\circ$ video and first-order ambisonic audio.% which we co-designed with a tourism operator in \note{location redacted for blind review}.

To allow for a more controlled comparison and reduce potential confounding factors the ``live'' experience was actually prerecorded, meaning that all conditions were based on identical video and audio recordings. This ensures that all participants had an identical experience, which we believe was necessary to improve the replicability of this study and make for a fairer comparison between conditions. In particular, early pilot testing revealed that the mobile connection on the lagoon was unstable, resulting in inconsistent video and audio quality which, while more faithful, resulted in an inconsistent experience between participants. This also allowed us to control other factors such as weather, sun glare, and wildlife which are famously unpredictable in the West Coast of New Zealand. The video was edited to blur the actual kayaker out in all conditions to mask the fact that they weren't the one talking.

The virtual tour, including prerecorded narration, is available to watch online as a $360^\circ$ video\footnote{\url{https://youtu.be/3CCf287JRos}}. We recommend watching this in a head-mounted display with headphones for the full experience.

This research was approved by the University of Otago Human Ethics Committee. Our study design and hypotheses were preregistered with the Open Science Foundation (OSF)\footnote{\url{https://osf.io/795bz}}.

\subsection{Conditions} 
We were interested in investigating the effects of two separate factors on virtual tourism experiences: whether the experience was live or prerecorded, and whether the experience included narration from a tour guide. We thus applied a multi-factor study design with two main independent variables: ``real-time'' streaming (yes|no) and narration (yes|no), as shown in \autoref{tab:study-conditions}. Our four study conditions were thus as follows:

\begin{itemize}
    \item \textbf{Non-Real-Time, No-Narrative (NRT-NN)}: Participants watched the $360^\circ$ video with ambisonic audio and were aware that it was prerecorded. No information, narration or audio was provided other than the ambient noise of the lagoon.
    \item \textbf{Non-Real-Time, With-Narrative (NRT-WN)}: Participants were aware that the experience was prerecorded. They listened to a narrative prerecorded by the paid actor (the ``tour guide'') during the experience.
    \item \textbf{Real-Time, No-Narrative (RT-NN)}: Participants in this deceptive condition were instructed their tour was happening live (a livestream); however, the same video and ambient audio were used as in the Non-Real-Time conditions. No narrative or additional information was provided.
    \item \textbf{Real-Time, With-Narrative (RT-WN)}: Participants in this deceptive condition were told they were watching a live stream, but actually experienced the same video and ambient audio used in the Non-Real-Time conditions. A narrative, performed by a paid actor posing as a tour guide, was provided. The actor pretended to be operating the kayak in the video and provided live narration via two-way Voice over IP (VoIP), facilitating conversation and questions from the participant.

\end{itemize}

\begin{table}[ht!]
    \begin{center}
        \begin{tabular}{l | l | l}
             & Narrative & No-Narrative \\
             \hline
             Real-Time & RT-WN & RT-NN \\
             Non-Real-Time & NRT-WN & NRT-NN
        \end{tabular}
    \end{center}
    \caption{Our four study conditions using a 2x2 design derived from two independent variables: ``Real-time'' streaming  (yes|no) and the presence of a voice-over narrative (yes|no).}
    \label{tab:study-conditions}
\end{table}

The information presented in the narrative was consistent for all participants in the NRT-WN condition as the audio was prerecorded, however the live aspect of the RT-WN condition made the consistency of information slightly more difficult to control as participants were free to ask the guide questions. An actor rehearsed the script while watching the video to prepare for live study sessions and recorded a rendition that played during the NRT-WN condition. An imaginary persona and backstory for the guide was also created in case participants asked any personal questions.

Prior to data collection, each participant was randomly assigned to a condition, with an equal number of participants per condition. Random assignment was maintained with the exception of the RT-* conditions when the deception was not possible to uphold. This occurred in two scenarios: 
\begin{enumerate}
    \item The participant session was scheduled after 3:30PM (within 2 hours of sunset) which would make lighting conditions inconsistent;
    \item The voice actor who performed for RT-WN was unavailable.
\end{enumerate}
In these situations, the condition number was simply switched with the next in line. 

\subsection{Hypotheses}
A recent review of live-streaming telepresence literature suggests that live-streaming techniques (eg. smartphone and 360-degree cameras) can provide the user with a strong sense of presence~\cite{Pfeil2021}. If the goal of a tourism experience is to \textit{transport} a person to a place, telepresence is a powerful tool. However, if the goal of an experience is to \textit{bond} a person to a place, presence cannot facilitate that alone. According to place attachment literature, mediated environments need to tell compelling stories in order to create meaningful VRTEs ~\cite{Chen2021}.
The effect of valence (whether an experience provokes positive or negative affect) and arousal on memory is an ongoing topic of research~\cite{Mancuso2023}. Yet few VR studies have explored memory biases, such as recall bias. As a result, we base our predictions on previous work investigating recall bias in non-VR studies, which observes a link between positive affect and overestimated memory~\cite{Mitchell1997, Colombo2020}. 

Based on prior research we had the following hypotheses:
\begin{itemize}
    \item \textbf{H1:} Those assigned to the Real-Time conditions (RT-*) will report higher levels of presence than in the Non-Real-Time conditions (NRT-*).
    \item \textbf{H2:} Those assigned to the With-Narrative conditions (*-WN) will exhibit higher levels of place attachment to the destination than in the No-Narrative conditions (*-NN).
    \item \textbf{H3:} There will be a significant positive correlation between Affect Grid scores and memory task scores.
\end{itemize}

%In accordance with these results we expect...

%However, the review also highlights that true presence, of the feeling of being there, cannoy be achieved without social ;"The foundational goal of telepresence is to grant the feeling of “being there,” but the H2H telepresence paradigm transcends sensory stimulation, and thus this goal cannot be achieved unless social needs of the Streamer and the Viewer are also met."

\subsection{Dependent Variables}\label{subsec:dependent-variables}
We measured the following dependent variables:

\begin{figure}
    \centering
    \includegraphics[width=.4\textwidth]{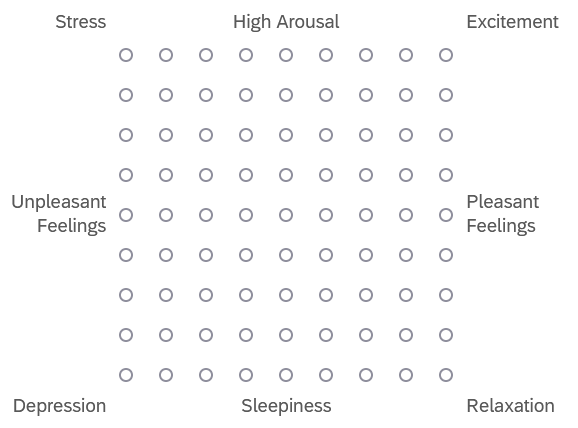}
    \caption{The Affect Grid, which we administered before and after the virtual kayaking experience. The grid consists of separate scales for Pleasure-Displeasure and Arousal-Sleepiness and tasks users with picking which point on the grid best describes their current affect.}
    \Description{A 9x9 grid of squares with eight labels around its outer edge; one on each corner, and one on each edge. These are, in clockwise order from the top left: Stress, High Arousal, Excitement, Pleasant Feelings, Relaxation, Sleepiness, Despression, and Unpleasant Feelings. This gives the grid two axes: (High Arousal to Sleepiness and Unpleasant Feelings to Pleasant Feelings)}
    \label{fig:affect_grid}
\end{figure}

\begin{itemize}
    \item \textbf{Affect:} Each participant's affect before and after the experience was measured using two instances of The Affect Grid~\cite{russell1989affect}; see \autoref{fig:affect_grid}.% represented as a 9x9 grid format (see \autoref{fig:affect_grid} but analysed as two separate 9-point Likert like scales for pleasure-displeasure and arousal-sleepiness dimensions.
    \item \textbf{Spatial Presence:} Participants' perceived spatial presence was measured after the experience using the Igroup Presence Questionnaire (IPQ)~\cite{Schubert2001, Tran2024}.
    \item \textbf{Place Attachment:} Place attachment was measured after the experience using the Abbreviated Place Attachment Scale (APAS)~\cite{Williams1992, Williams2003}; but we also included the original items related to nature bonding due to the ecological focus of the experience. %one feels with the destination environment, has been a topic of interest in environmental psychology in recent decades. Through extensive study, multiple subdimensions have emerged, most notably place identity, place dependence, and nature bonding (where appropriate).
    \item \textbf{Tourism experience:} The four realms of tourism experiences were measured after the experience using a questionnaire proposed by Pine and Gilmore~\cite{Pine1998}. These realms are ``entertainment'', ``education'', ``esthetics,'' and ``escapism''. 
    \item \textbf{Simulator sickness:} Potential symptoms of simulator sickness were measured before and after the experience using the Cybersickness in Virtual Reality Questionnaire (CSQ-VR)~\cite{Kourtesis2023}.
    \item \textbf{Memory task}: A screenshot was taken from the experience and manipulated to make it look better (eg. higher saturation, clearer skies and water, more wildlife) or worse (greyer skies, darker water, etc.). Five versions of the image were presented to each participant: one real image, two ``positive'' images, and two ``inferior'' images. Participants were instructed to select the image that most accurately reflected their memory of the experience. This measure was inspired by previous research on memory bias and ``rosy retrospection''~\cite{Panagiotopoulou2022}.
\end{itemize}

We also conducted a semi-structured interview after the experience for additional feedback. Participant responses were not formally analysed (eg. thematic analysis) but were used to screen for deception failures, find justifications for outliers in the quantitative data, and guide future research.

\subsection{Study Material and Apparatus}
The 360$^\circ$ video used for the experience was recorded using an Insta360 ONE X2\footnote{\url{https://www.insta360.com/product/insta360-onex2}} at a resolution of 5760x2880. Ambient audio was recorded with a ZOOM H2n\footnote{\url{https://zoomcorp.com/en/us/handheld-recorders/handheld-recorders/h2n-handy-recorder/}} and rendered using the first-order ambisonic B-format, allowing sound to be spatially situated and react to the participant's head movements.% Both devices were placed at the front of the kayak using a custom 3D-printed mounting bracket to give the impression that the participant was sharing the kayak with the tour guide (see \autoref{fig:teaser}).

%We captured roughly fourteen takes of the tour and, after review, decided on a take that is 14:52 long. The main deciding points were mainly finding suitable lighting conditions for the $360^\circ$ camera, abundant bird life, and minimising other issues such as glare and water droplets on the lens. However, most of the takes would have qualified and the differences between them are minimal.

\begin{figure}
    \centering
    \includegraphics[width=\linewidth]{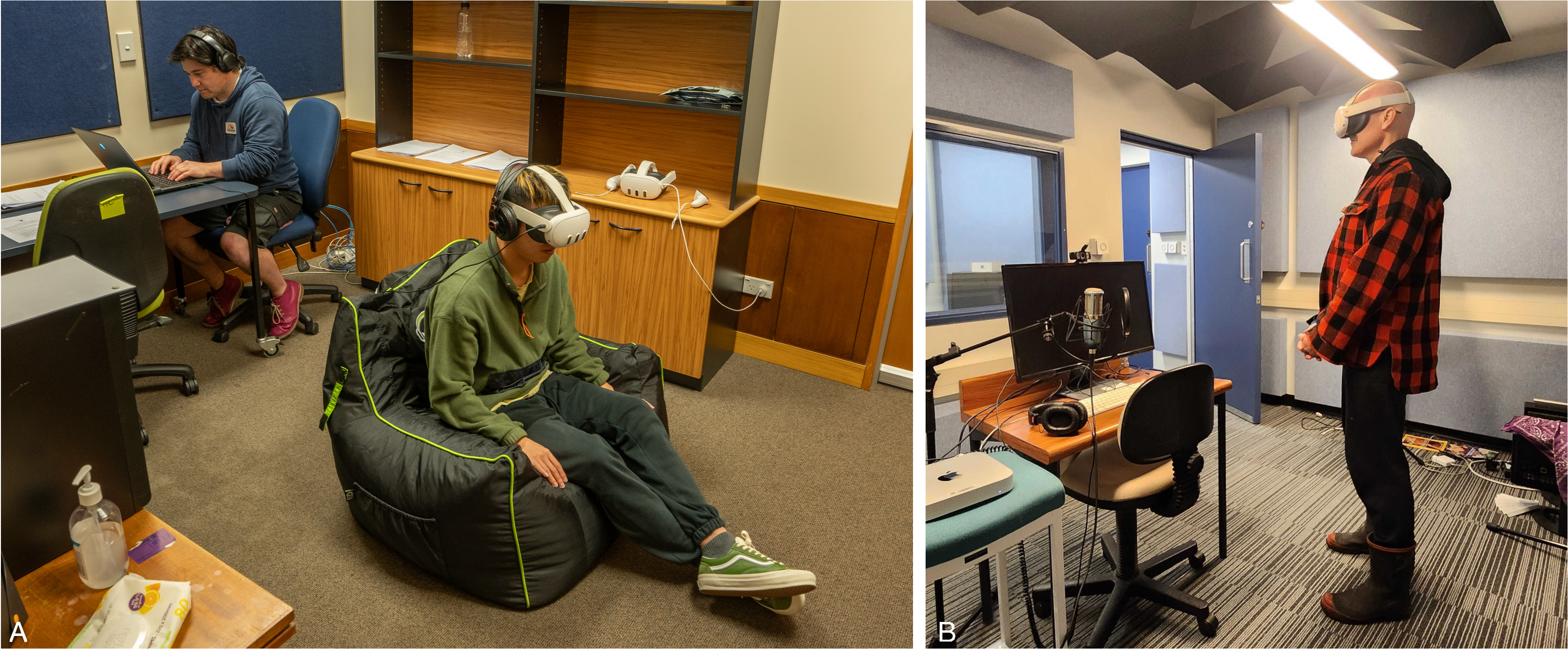}
    \caption{The setup used for our user study. A) Participants were seated on a low beanbag roughly the same height as the $360^\circ$ camera to give the illusion they were seated in a kayak. The experience was presented in a Meta Quest 3 with semi-open back headphones connected to play the ambisonic audio. The study media was also present and monitoring the experience through a laptop. B) In the RT-WN condition, a paid actor pretended to pilot the kayak but was actually in a sound booth with their own Meta Quest 3 and providing narrative via VoIP.}
    \Description{Two images horizontally next to each other. The left image shows a study participant seated in a bean bag and wearing a white virtual reality headset while wearing a Meta Quest 3; the study mediator is seated at a laptop behind them and wearing a pair of headphones while monitoring the VR experience. The right image shows the paid actor standing within a sound booth and wearing a Meta Quest 3; it is implied that they are talking to the study participant.}
    \label{fig:StudySetup}
\end{figure}

The ambisonic audio and 360$^\circ$ video recordings were combined in the Unity game engine and played back on a Meta Quest 3. The audio in the two narrative conditions (*-WN) was spatially situated behind the player camera and EQ matched with the ambient recording to give the illusion that it came from the person piloting the kayak. In the ``Real-time, With-Narrative'' (RT-WN) condition, the narration was livestreamed from a paid actor using a separate Meta Quest 3 in an audio-controlled room; the connection between participant and narrator was facilitated using the Normcore API\footnote{https://normcore.io/}, with video playback synchronised across the network to ensure that both parties saw the same stage of the experience. The audio in the ``Non-Real-Time, With-Narrative'' (NRT-WN) condition was prerecorded by the same actor and played back in sync with the video. The full study setup can be seen in \autoref{fig:StudySetup}.

\begin{figure*}[t]
    \centering
    \includegraphics[width=\textwidth]{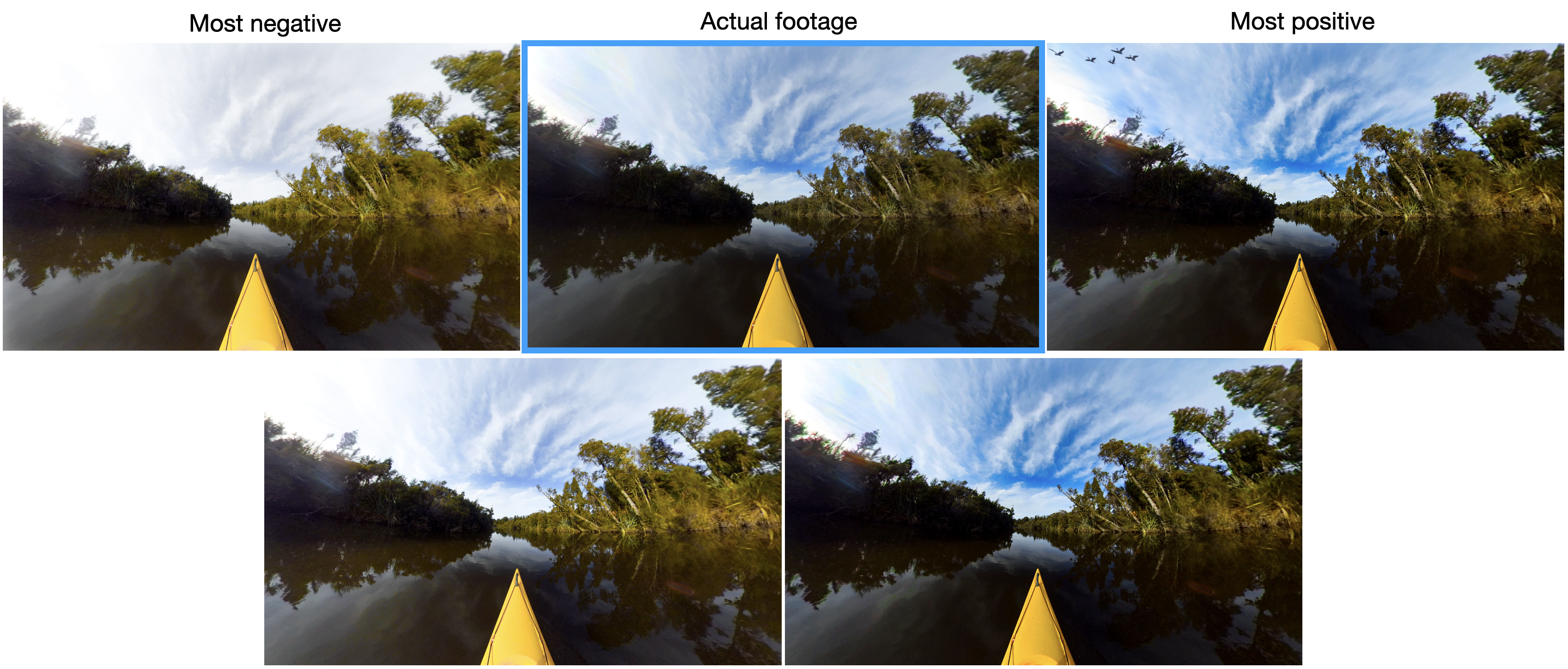}
    \caption{The five images participants could choose from to visually represent their memory of the experience. One was an actual screenshot, two were edited to look worse (less saturated, etc.) and two were edited to look visually better (higher saturation, more wildlife, etc.).}
    \Description{Five images looking out at a lagoon from the front of a kayak. They are arranged in order from "worst" to "best"; the worse images are generally greyer and appear more overcast, while the better images are brighter, less cloudy, and appear to have clearer water. The rightmost image, the ``best'' one, also has a flock of birds edited into its top-left corner.}
    \label{fig:memory_task}
\end{figure*}

\subsection{Procedure}
\textbf{Onboarding}: After welcoming each participant, they were given an information sheet explaining that the study's purpose is to test the effects of a short VR tourism experience and provide feedback to guide future development. We followed up by explicitly stating the risk of simulator sickness, how to recognise it, and what to do in the event of symptom onset; any participant who felt ill was to be immediately withdrawn from the study, though this was not necessary for any participant. At this point participants also signed their Consent to Participate.

\textbf{Pre-Experience Questionnaires}: Before starting the study we asked each participant to fill out a demographics form. If participants indicated that they had a history of adverse effects from VR or motion sickness they were excluded from the study. Participants also completed the CSQ-VR~\cite{Kourtesis2023}) and Affect Grid~\cite{russell1989affect} to get baseline values for their current simulator sickness symptoms and affect respectively. This form and all subsequent questionnaires were administered on a standard desktop display using Qualtrics$^\text{XM}$.

\textbf{Briefing}: Participants were seated in a designated bean bag chair, as seen in \autoref{fig:StudySetup} and informed that they would watch a roughly 15-minute guided tour of Ōkārito lagoon in VR and be asked about their experience afterwards. Participants in the Non-Real-Time conditions (NRT-*) were told that the experience was prerecorded, but participants in the Real-Time conditions (RT-*) were deceptively told that the video and audio they were about to see was being streamed live from a tour guide currently on the lagoon.

\textbf{Experimental Task}: Participants were fitted with the Meta Quest 3 pre-loaded with the experience and a pair of semi-open-back over-ear headphones. Once the participant verbally confirmed they were ready, the study coordinator remotely started the experience from their laptop computer; in the two Real-Time conditions (RT-*) they pretended to coordinate this with the tour guide to sell the illusion that the experience was happening live. The participant then watched the experience. However, those in the RT-WN condition were able to talk to the ``tour guide'' via VoIP. The study coordinator stayed in the room throughout the experience for health and safety reasons and could monitor the experience in real-time on a separate laptop.

\textbf{Post-Experience Questionnaires and Interview}: Immediately following the experience, the participants were asked to complete the post-experience questionnaires. Participants were also interviewed face-to-face to gather qualitative feedback about the experience that may not have been captured by the questionnaires.

%
%These were, in order 
%Affect Grid~\cite{russell1989affect}
%    \item The Cybersickness Questionnaire (CSQ-VR)~\cite{Kourtesis2023}
%    \item The iGroup Presence Questionnaire (IPQ)~\cite{Schubert2001}
%    \item The Abbreviated Place Attachment Scale (APAS)~\cite{Williams2003}
%    \item The Realms of Tourism questionnaire~\cite{Pine1998}
%    \item The memory task described in \autoref{subsec:dependent-variables}
%\end{itemize}

\textbf{Debrief}: Participants were thanked and compensated for their time with a \$20 supermarket voucher. Those in the two Real-Time conditions (RT-*) were notified at this point that they had been deceived during the experience and that the tour had been prerecorded; if it was clear that a participant was already aware of the deception their data was discarded to prevent this impacting the study results.

\section{Results}
93 participants were recruited from the community between the ages of 18 and 74 ($M = 29.13$, $SD = 11.41$). 13 were excluded from analysis due to technological failures (e.g. HMD popups, incomplete surveys) or because they saw through the deception, leaving 80 participants in the final analysis (20 per condition)\footnote{Our study design was pre-registered with 52 participants but this was increased on reviewer request.}. 30 were male, 47 were female, two were non-binary/gender diverse, and one preferred not to disclose their gender. None of the participants had a history of severe motion sickness, and all had normal or corrected-to-normal vision. 26 participants had never tried VR before, while the rest had at least some experience with the technology. 

\begin{figure*}[t!]
    \centering
    \includegraphics[width=\textwidth]{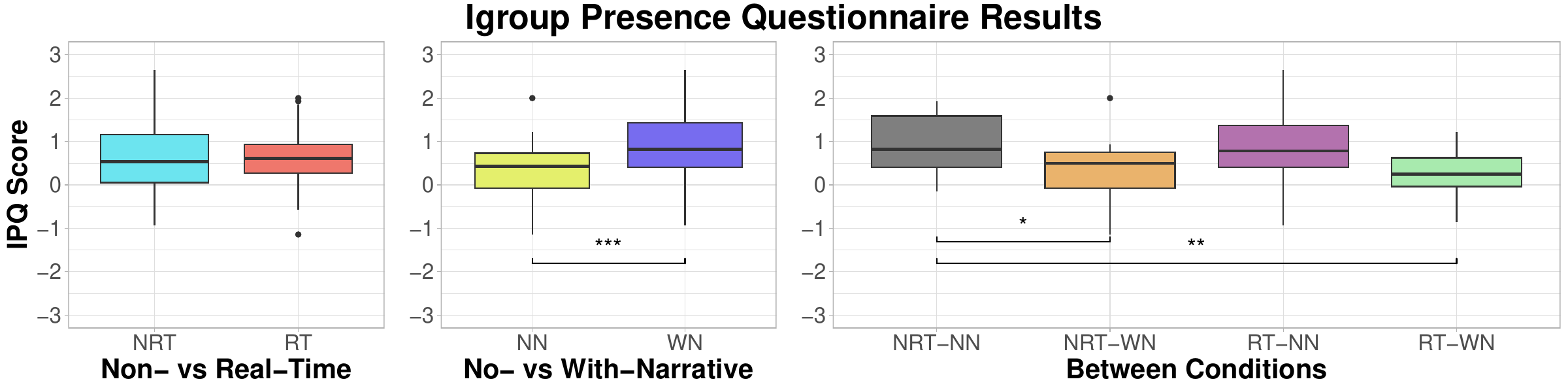}
    \caption{IPQ questionnaire results for (Left): Non-Real-Time vs Real-Time groups, (Centre): No-Narrative vs With-Narrative groups, (Right): By condition. * = significant difference.}
    \Description{Three box plots comparing the distributions of IPQ scores between groups. The leftmost image compares the Non-Real-Time to the Real-Time group; the distributions are roughly equal and centred at around 0.6. The centre image compares the No-Narrative to the With-Narrative group; the With-Narrative distribution is slightly higher (median = 0.82) than the No-Narrative group (median = 0.43), and a bar with an asterisk above it shows that the difference between them is statistically significant. The rightmost image shows the distributions of the four individual conditions: NRT-NN (median = 0.25), NRT-WN (median = 0.786), RT-NN (median = 0.50), and RT-WN (median = 0.82). There are no bars drawn between the distributions, indicating that none of these differences are statistically significant.}
    \label{fig:results/ipq}
\end{figure*}

\begin{table*}[t!]
    \centering
    \begin{tabular}{ccccccccccc} 
        \toprule
         & \multicolumn{2}{c}{IPQ Score} & \multicolumn{2}{c}{General Presence} & \multicolumn{2}{c}{Spatial Presence} & \multicolumn{2}{c}{Involvement} & \multicolumn{2}{c}{Realism} \\
         \cmidrule(lr){2-3}\cmidrule(lr){4-5}\cmidrule(lr){6-7}\cmidrule(lr){8-9}\cmidrule(lr){10-11}
         & $M \pm IQR$ & $p$ & $M \pm IQR$ & $p$ & $M \pm IQR$ & $p$ & $M \pm IQR$ & $p$ & $M \pm IQR$ & $p$ \\
         \midrule
         NRT-* & $0.53 \pm 1.11$ & & $1.00 \pm 1.00$ & & $0.90 \pm 1.10$ & & $0.63 \pm 1.50$ & & $0.00 \pm 0.81$ & \\
         RT-* & $0.61 \pm 0.66$ & \multirow{-2}{*}{.58} & $0.80 \pm 1.05$ & \multirow{-2}{*}{.15} & $0.96 \pm 1.02$ & \multirow{-2}{*}{.84} & $0.75 \pm 1.81$ & \multirow{-2}{*}{.81} & $0.00 \pm 1.00$ & \multirow{-2}{*}{.84} \\
         \midrule
         *-NN & \cellcolor{sigcell} $0.43 \pm 0.80$ & \cellcolor{sigcell} & \cellcolor{sigcell}$1.00 \pm 1.25$ & \cellcolor{sigcell} & \cellcolor{sigcell}$0.80 \pm 0.70$ & \cellcolor{sigcell} & \cellcolor{sigcell} $0.25 \pm 1.38$ & \cellcolor{sigcell} & $0.00 \pm 0.81$ & \\
         *-WN & \cellcolor{sigcell} $0.82 \pm 1.02$ & \multirow{-2}{*}{\cellcolor{sigcell}$.0006$} & \cellcolor{sigcell}$2.00 \pm 2.00$ & \multirow{-2}{*}{\cellcolor{sigcell}.01} & \cellcolor{sigcell}$1.00 \pm 1.40$ & \multirow{-2}{*}{\cellcolor{sigcell}.02} & \cellcolor{sigcell}$1.00 \pm 1.56$ & \multirow{-2}{*}{\cellcolor{sigcell}$.005$} & $0.13 \pm 1.00$ & \multirow{-2}{*}{.10} \\
         \bottomrule
    \end{tabular}
    \caption{Median $M$, Inter-Quartile Range $IQR$, and two-way ANOVA results $p$ from the Igroup Presence Questionnaire and its subscales per group. Each questionnaire item was scored in the range [-3,3] with higher scores indicating a greater degree of presence. Statistically significant differences are highlighted in yellow.}
    \label{tab:IPQ}
\end{table*}

%\subsection{Quantitative Analysis}\label{sec:results/quant}
All data was tested for normality using the Shapiro-Wilkes test. We treated Likert scale responses from the IPQ, Realms of Tourism, Place Attachment, and Affect Grid questionnaires as non-ordinal due to their subjective nature. Comparisons between groups (eg. With-Narrative vs No-Narrative) were performed using two-way ANOVA ($\alpha=0.05$); we first used the Aligned Rank Transform (ART) to address violations of normality and heterogeneity of variances\footnote{Our pre-registered study design stated that Wilcoxon Rank Sum tests would be used but this was changed to ART and ANOVA on reviewer request.}. Comparisons between the four conditions were performed with Kruskal-Wallis tests, with post-hoc comparisons using Dunn's test adjusted with the Holm-Bonferroni method.

\subsection{Spatial Presence}
Our first hypothesis was that spatial presence would be rated higher in the Real-Time conditions (RT-*) than in the Non-Real-Time conditions (NRT-*). The mean of each participant's fourteen IPQ responses was calculated and used as an overall IPQ score which was found to be normally distributed. Each item was first scaled from -3 to 3 based on advice from a recent meta review~\cite{Tran2024}. A two-way ANOVA on the ART-transformed data did not find a significant interaction between narrative and livestreaming ($p = .98$) and so the two variables were analysed separately. See \autoref{fig:results/ipq} for a graphical representation of our results or \autoref{tab:IPQ} for a full statistical analysis.

A two-way ANOVA did not find a significant difference in IPQ scores between the RT-* and NRT-* ($p = .58$), and  Kruskal-Wallis test did not find significant differences between individual conditions ($p = .52$). \textbf{H1} is thus not supported.

\begin{figure*}[t]
    \centering
    \includegraphics[width=0.32\linewidth]{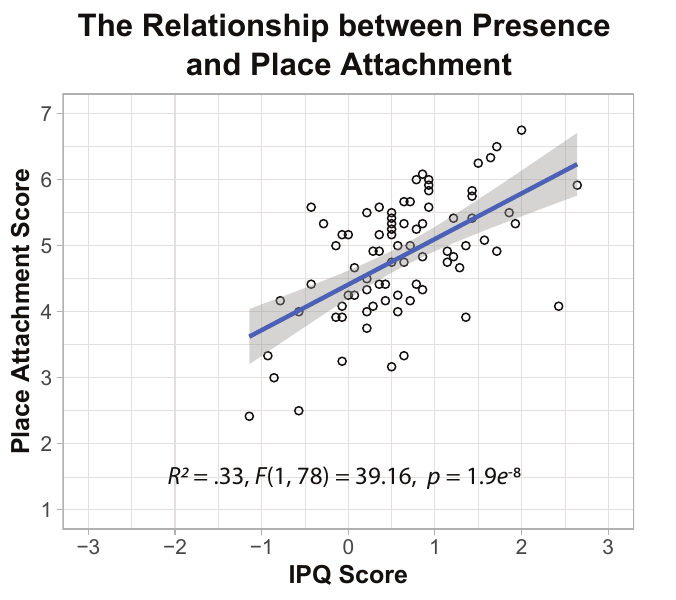}
    \includegraphics[width=0.32\linewidth]{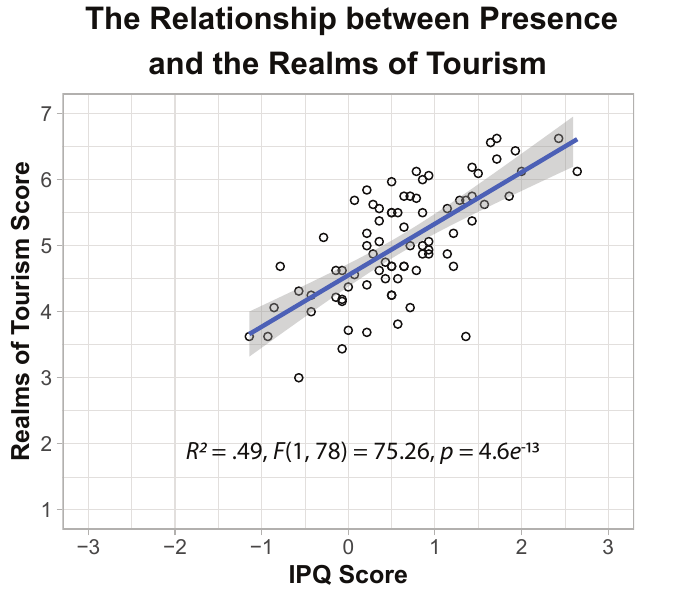}
    \includegraphics[width=0.32\linewidth]{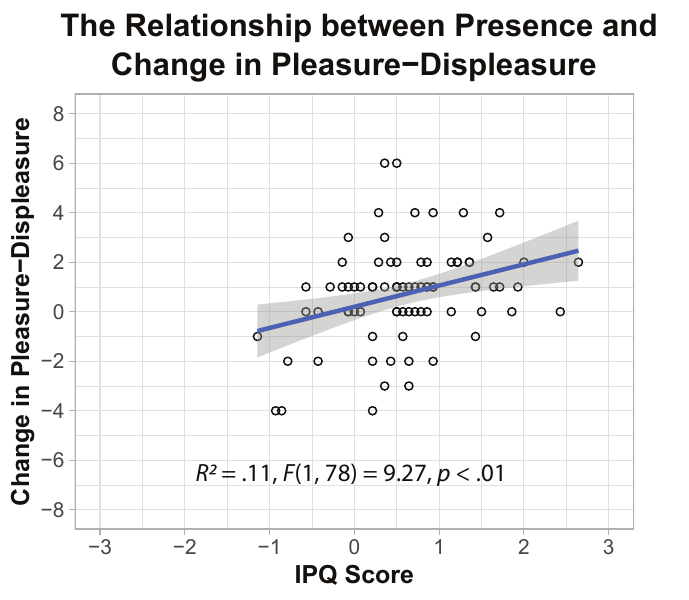}
    \caption{The relationship between spatial presence and Place Attachment (left), the Realms of Tourism (centre), and participants' change in Pleasure-Displeasure (right).}
    \Description{Three scatter plots showing the correlation between presence and place attachment (left), presence and the realms of tourism (right), and presence and the change in Pleasure-Displeasure (right). All three graphs show a tight correlation, with a trend line at roughly a 45 degree angle.}
    \label{fig:results/lm}
\end{figure*}

Though not one of our original hypotheses, a two-way ANOVA found a significant difference in IPQ scores ($p < .001$, $\eta^{2} = 0.14$, post-hoc power $PPA = .95$) between the With-Narrative *-WN (median $M = 0.82$, interquartile range $IQR = 0.71$) and No-Narrative *-NN ($M = 0.43$, $IQR = 0.80$) groups; the *-WN conditions were rated ``High'' in presence compared to existing studies while the *-NN conditions were rated as only ``Moderate''~\cite{Tran2024}. Looking at individual subscales, two-way ANOVA found significant differences between the *-WN and *-NN groups for general presence ($p = .01$, $\eta^2 = 0.08$, $PPA = .73$), spatial presence ($p = .02$, $\eta^{2} = 0.07$, $PPA = .70$), and involvement ($p < .01$, $\eta^{2} = 0.10$, $PPA = .84$), but not for realness ($p = .10$).

Simple linear regression was used to determine if there was a link between induced presence and our other measurements. We found a statistically significant relationship between presence and place attachment ($R^2 = .33$, $F(1, 78) = 39.16$, $p = 1.9e^{-8}$) with the model $PA = -1.73 + 0.48 \times IPQ$. We also found a significant relationship between presence and the realms of tourism score ($R^2 = .49$, $F(1, 78) = 75.26$, $p < 4.6e^{-13}$) with the model $ROT = -2.56 + 0.63 \times IPQ$, and between presence and change in pleasure-displeasure ($R^2 = .11$, $F(1, 78) = 9.27$, $p < .01$) with the model $\Delta_{pl-dpl} = 0.52 + 0.12 \times IPQ$. See \autoref{fig:results/lm} for a graphical representation of these relationships.

%Kruskal-Wallis tests also found no significant difference between individual conditions for overall IPQ score ($p = .09$), General Presence ($p = .33$), Spatial Presence ($p = .68$), Involvement ($p = .07$), or Realism ($p = .43$). 

%%change the INV x Narrative ANOVA score (above, 0.012) to a t-test value; Kruskal Wallace test? Do a Shapiro-wilk for INV/subscales 
%We performed an exploratory analysis on the IPQ items by group using an ANOVA test. We found that responses to item no. 10 (“I was completely captivated by the virtual world.”) had a significant relationship to group (p = 0.024). 

%% do we want to add the t-test between Item #10 and narrative?

\subsection{Place Attachment}
\begin{figure*}[t!]
    \centering
    \includegraphics[width=\textwidth]{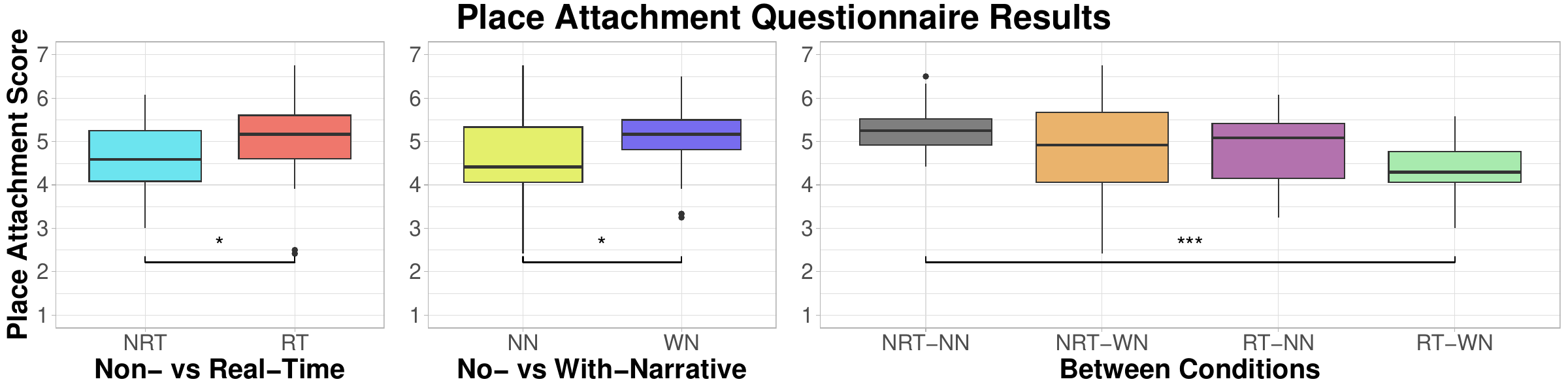}
    \caption{Results from the Place Attachment questionnaire, from left to right: (1): Non-Real-Time vs Real-Time conditions, (2): No-Narrative versus With-Narrative conditions, (3): By condition. * = significant difference.}
    \Description{Three box plots comparing the distributions of place attachment scores between groups. The leftmost image compares the Non-Real-Time (median 4.58) to the Real-Time group (median = 5.17), with a bar between them indicating their difference is significant. The centre image compares the No-Narrative to the With-Narrative group; the With-Narrative distribution is slightly higher (median = 5.17) than the No-Narrative group (median = 4.42), again with a bar to show significance. The rightmost image shows the distributions of the four individual conditions: NRT-NN (median = 4.29), NRT-WN (median = 5.08), RT-NN (median = 4.92), and RT-WN (median = 5.25).}
    \label{fig:results/pa}
\end{figure*}

Our second hypothesis was that participants assigned to the With-Narrative conditions (*-WN) would exhibit significantly higher place attachment to the destination than those in the No-Narrative conditions (*-NN). The mean of all Place Attachment items was calculated per participant to obtain a Place Attachment ($PA$) score from 1-7, which a Shapiro-Wilkes test found to be normally distributed. A two-way ANOVA on the ART-transformed data did not find a significant interaction between narrative and livestreaming ($p = .95$) and so the two variables were analysed separately. See \autoref{fig:results/pa} for a graphical representation of these results, or \autoref{tab:PA} for a full statistical analysis.

A two-way ANOVA found a significant difference in Place Attachment scores ($p = .01$, $\eta^2 = 0.08$, $PPA = .73$) between the *-WN ($M = 5.17$, $IQR = 0.69$) and *-NN ($M = 4.42$, $IQR = 1.27$) groups, supporting our second hypothesis \textbf{H2}. A significant difference was also found between the two groups for the Place Identity subscale ($p = .02$, $\eta^2 = .15$, $PPA = .96$), but not for Nature Bonding ($p = .08$) or Place Dependence ($p = .22$)

A two-way ANOVA also found a significant difference in Place Attachment ($p = .02$, $\eta^2 = 0.07$, $PPA = .70$) between the RT-* ($M = 5.17$, $IQR = 1.00$) and NRT-* ($M = 4.58$, $IQR = 1.17$) groups. A significant difference was also found between these groups for the Nature Bonding ($p = .01$, $\eta^2 = .09$, $PPA = .77$) and Place Identity ($p = .04$, $\eta^2 = .05$, $PPA = .55$) subscales, but not for Place Dependence ($p = .12$).

Simple linear regression found a significant relationship between Place Attachment and the Realms of Tourism score ($R^2 = .54$, $F(1, 78) = 89.77$, $p = 1.2e^{-15}$ with the model $PA = 0.89 + 0.78 \times RoT$. A significant relationship was also found between Place Attachment and a change in Pleasure-Displeasure ($R^2 = .09$, $F(1, 78) = 8.11$, $p < .01$) with the model $\Delta_{pl-dpl} = 4.73 + 0.14 \times PA$.

\begin{table*}[t!]
    \centering
    \begin{tabular}{ccccccccc} 
        \toprule
         &\multicolumn{2}{c}{Place Attachment}&\multicolumn{2}{c}{Nature Bonding}&\multicolumn{2}{c}{Place Identity}&\multicolumn{2}{c}{Place Dependence} \\
         \cmidrule(lr){2-3}\cmidrule(lr){4-5}\cmidrule(lr){6-7}\cmidrule(lr){8-9}
         & $M \pm IQR$ & $p$ & $M \pm IQR$ & $p$ & $M \pm IQR$ & $p$ & $M \pm IQR$ & $p$ \\
         \midrule
         NRT-* & \cellcolor{sigcell}$4.58 \pm 1.17$&                       \cellcolor{sigcell}& \cellcolor{sigcell}$5.25 \pm 1.19$&                       \cellcolor{sigcell}& \cellcolor{sigcell}$5.00 \pm 1.12$&                       \cellcolor{sigcell}& $3.88 \pm 1.44$&                        \\
         RT-*  & \cellcolor{sigcell}$5.17 \pm 1.00$& \multirow{-2}{*}{\cellcolor{sigcell}.01}& \cellcolor{sigcell}$5.75 \pm 1.00$& \multirow{-2}{*}{\cellcolor{sigcell}.01}& \cellcolor{sigcell}$5.50 \pm 1.00$& \multirow{-2}{*}{\cellcolor{sigcell}.04}& $4.25 \pm 1.31$& \multirow{-2}{*}{.12}\\
         \midrule
         *-NN & \cellcolor{sigcell}$4.42 \pm 1.27$&                       \cellcolor{sigcell}& $5.34 \pm 0.94$ &                      & \cellcolor{sigcell} $4.75 \pm 1.31$& \cellcolor{sigcell}                     & $3.75 \pm 1.31$&                       \\
         *-WN & \cellcolor{sigcell}$5.17 \pm 0.69$& \multirow{-2}{*}{\cellcolor{sigcell}.01}& $5.64 \pm 1.02$ & \multirow{-2}{*}{.08}& \cellcolor{sigcell} $5.50 \pm 1.00$& \multirow{-2}{*}{\cellcolor{sigcell}.0004}& $4.25 \pm 1.31$& \multirow{-2}{*}{.23}\\
         \bottomrule
    \end{tabular}
    \caption{Median $M$, Inter-Quartile Range $IQR$, and two-way ANOVA results $p$ from the Place Attachment questionnaire and its subscales per group. Each questionnaire item was scored in the range [1,7] with higher results indicating more place attachment. Statistically significant differences are highlighted in yellow.}
    \label{tab:PA}
\end{table*}

%%H2: WN x PA ANOVA p = 0.89; replace with t-test 

%%There are a few possible explanations for these findings. Perhaps VR is an effective medium at facilitating PA, without the need for any additional information presented to the viewer. Additionally the PA scale was not created with VR content in mind and we are, as far as we know, the first study to use this scale in this context. 

%An ANOVA test found no significant differences in Total PA between Groups (p = 0.142). We performed an exploratory analysis on the sub scales for PA, revealing a trend towards significance between the Real-time condition and Nature Bonding sub scale (p = 0.084). Additionally, we saw a significant relationship between responses to PA Scale Item No.2 (“I would feel less attached to Ōkārito if the native plants and animals that live here disappeared.”) by Group, according to an ANOVA test (p = 0.04). 

%%can you say “trend” or “weak relationship” ? Also do we want to add the t-test between Item #2 and real-time? 

\subsection{Affect and Memory}
\begin{figure}[t!]
    \centering
    \includegraphics[width=\linewidth]{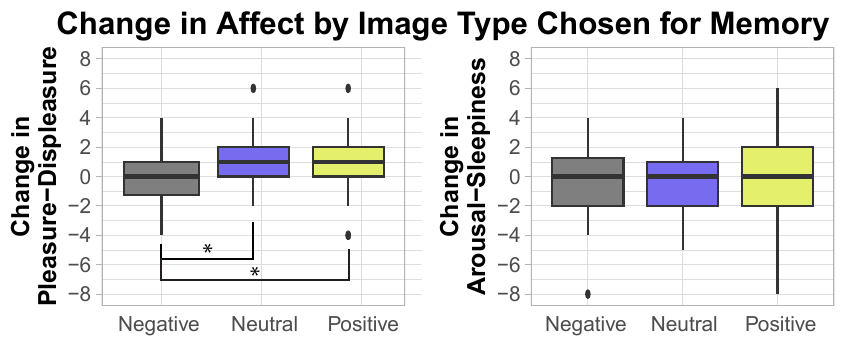}
    \caption{Participants' change in pleasure-displeasure (left) and arousal-sleepiness (right) during the experience grouped by which type of image they chose to represent their memory of it: negatively altered, faithful to the actual experience, or positively altered}
    \Description{Two box plots showing how many participants selected a negative, neutral, or positive image in the memory task, and the distribution in change in arousal-sleepiness (left graph) and pleasure-displeasure (right graph) for each image type. The left graph, indicating the change in pleasure-displeasure for each image type, shows Neutral and Positive centred at 1 and Negative sitting slightly lower at 0. The graph on the right, indicating the change in pleasure-displeasure for each image type, shows all three groups centred around 0.}
    \label{fig:affect-vs-memory}
\end{figure}

Our third hypothesis was that there would be a significant positive correlation between Affect Grid scores and memory task scores, indicating that a more positive affect would be associated with more positive memories of the experience. A participant's change in Affect (either Arousal-Sleepiness or Pleasure-Displeasure) was calculated as the difference between their pre- and post-experiment Affect Grid responses which we treated as two nine-point Likert scales. The median change in ``Pleasure-Displeasure'' during the experience was 1.00 ($IQR = 2.00$), while the median change in ``Arousal-Sleepiness'' was 0.00 ($IQR = 4.00$). Refer to \autoref{fig:affect-vs-memory} for a distribution of affect change grouped by the image chosen, \autoref{fig:results/affect_change} for the affect change per group, or \autoref{tab:Affect} for a full statistical analysis.

We performed multinomial logistic regression to examine participants' likelihood to select a negatively- or positively-altered image to describe their memory of the experience over an actual screenshot of it (or a ``Neutral'' image) depending on their change in ``pleasure-displeasure'' and ``arousal-sleepiness''. Images were classified as ``Neutral'' if they were a direct screenshot from the experience (selected by 21 participants), ``Negative'' if they had been altered to look worse than the screenshot (eg. by decreasing saturation or adding clouds to the sky) (selected by 16 participants), or ``Positive'' if they were altered to look better than the screenshot (eg. by increasing saturation or adding wildlife) (selected by 43 participants). See \autoref{fig:affect-vs-memory} for a graphical representation of these results.

A change in ``Pleasure-Displeasure'' was found to be a strong predictor for participants' choice of image, with an increase of one on the scale (towards Pleasure) decreasing their odds of selecting a Negative image over the Neutral one by 33.51\% ($p = .03$, 95\% CI [0.46, 0.96]). The model does not support that a change in Pleasure would also affect the likelihood of selecting a Positive image over the Neutral one ($p = .63$), however it does increase the odds of a Positive image being chosen over a Negative one by 40.51\% ($p = .04$, 95\% CI [1.00, 1.90]). The model also does not support that a change in Arousal-Sleepiness would affect participants' likelihood of selecting a Positive ($p = .96$) or Negative ($p = .75$) image over the Neutral one, or a Positive one over a Negative one ($p = .75$). Our third hypothesis is thus partially supported, though only for Pleasure-Displeasure and not for Arousal-Sleepiness.

The model also found that the odds of a participant selecting a Positive image is 119.38\% higher than selecting the Neutral image ($p = .01$, 95\% CI [1.20, 4.00]) or 143.72\% higher than selecting a Negative image ($p < .01$, 95\% CI [1.30, 4.40]) when the change in Pleasure-Displeasure and Arousal-Sleepiness are both zero.

\subsection{Change in Affect}
\begin{figure*}[t!]
    \centering
    \includegraphics[width=\textwidth]{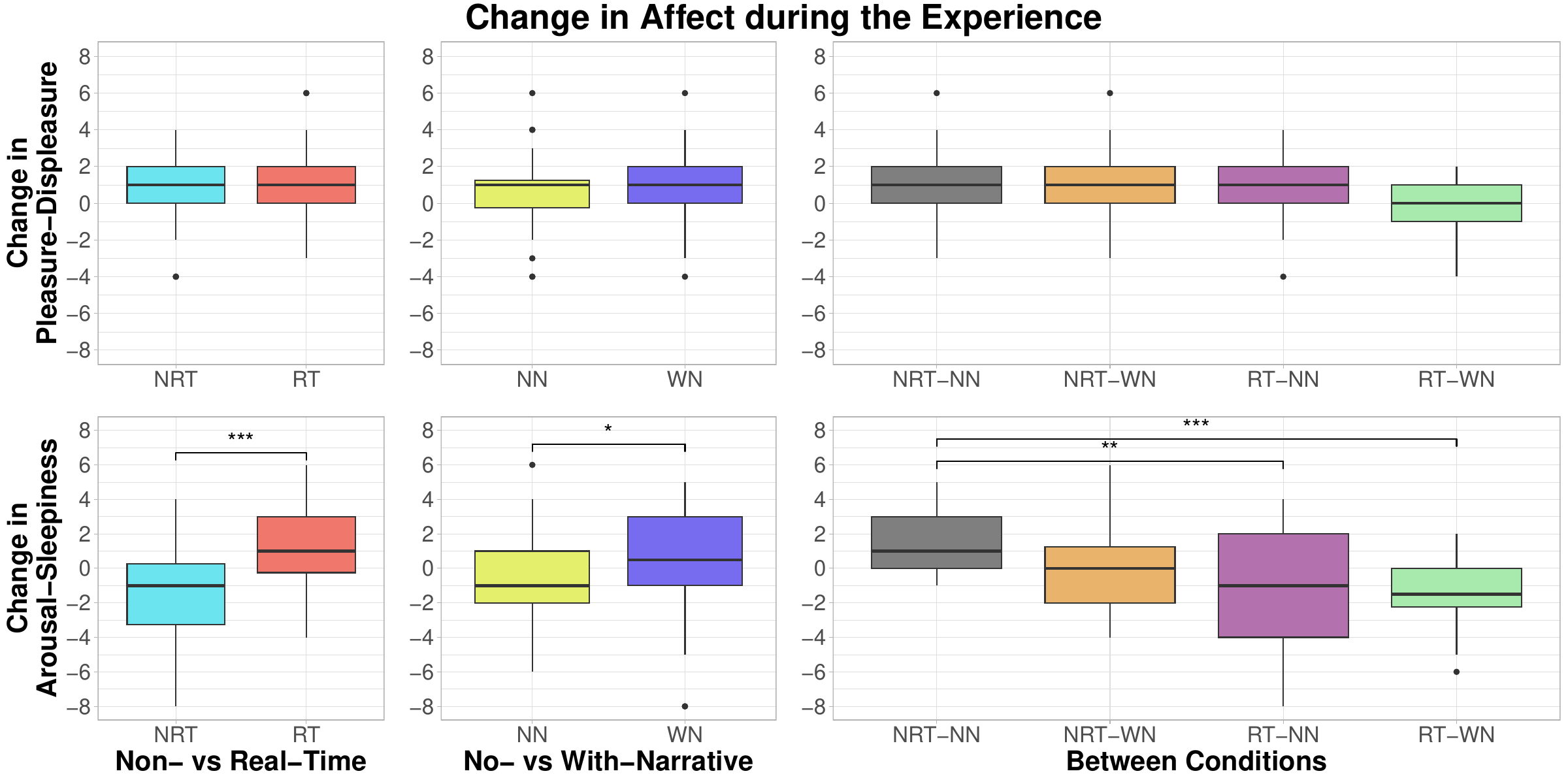}
    \caption{Participants' change in Pleasure-Displeasure (top row) and Arousal-Sleepiness (bottom row) during the experience, separated by Non-Real-Time vs Real-Time (left column), No-Narrative vs With-Narrative (centre column), and per condition (right column). * = significant difference.}
    \Description{Two rows of three box plots each to show the change in pleasure-displeasure (top row of graphs) and arousal-sleepiness (bottom row) between the Non-Real-Time group and Real-Time group (left column of graphs), between the No-Narrative and With-Narrative groups (centre column), and between conditions (right column). All groups in the top row are centred around 1, indicating no difference and a slight trend towards pleasure. The bottom row of graphs, representing Arousal-Sleepiness, shows more differences. NRT (median -1) is significantly lower than RT (median 1), NN (median -1) is significantly lower than WN (median 0.5), and NRT-NN (median -1.5) is significantly lower than both RT-NN (median 0) and RT-WN (median 1).}
    \label{fig:results/affect_change}
\end{figure*}

The change in participants' Affect as a result of the experience was also compared between groups. A two-way ANOVA on the ART-transformed data did not find a significant interaction between narrative and livestreaming for either the Arousal-Sleepiness scale ($p = .22$) or for the Pleasure-Displeasure scale ($p = .46$). See \autoref{fig:results/affect_change} for a graphical representation of these results, or \autoref{tab:Affect} for a full statistical analysis.

Two-way ANOVA found a significant difference in the change in Arousal-Sleepiness between the RT-* and NRT-* groups ($p < .001$, $\eta^2 = .16$, $PPA = .97$), with participants in the NRT-* group trending towards Sleepiness ($M = -1.00$, $IQR = 3.50$) and those in the RT-* group trending towards Arousal ($M = 1.00$, $IQR = 3.25$).

Two-way ANOVA did not find a significant difference in the change in Pleasure-Displeasure between the RT-* and NRT-* groups ($p = .26$) or between the *-WN and *-NN groups ($p = .19$). Similarly, two-way ANOVA failed to find a significant difference in the change in Arousal-Sleepiness between the *-WN and *-NN groups ($p = .07$).

%Examining individual conditions, a Kruskal-Wallis test found a significant difference in arousal-sleepiness between conditions ($p < .01$, $H = 0.23$, $df = 3$). Dunn's Test revealed significant differences in arousal-sleepiness between the RT-WN and NRT-WN conditions ($Z = 2.60$, $p < .05$) and between the RT-WN and NRT-NN conditions ($Z = 3.61$, $p < .01$), with participants in the RT-WN condition seeing a mean increase in arousal ($M = 2.23$, $SD = 1.92$) and participants in the NRT-WN ($M = -0.69$, $SD = 3.25$) and NRT-NN ($M = -1.62$, $SD = 2.22$) seeing a mean decrease in arousal (or increase in sleepiness).

%No significant difference was found in the change in pleasure-displeasure between conditions (Kruskal-Wallis, $p = .44$). Linear regression also showed no relationship between arousal-sleepiness and pleasure-displeasure ($p = .86$, $R^2 = .03$).

%An ANOVA test revealed a significant relationship between change in arousal between Groups (p = 0.00016). Furthermore, ANOVA tests revealed relationships between change in arousal and both Narrative (p = 0.065) and Real-time (p = 0.001) conditions, however only the RT condition relationship is significant.

\subsection{Realms of Tourism}
\begin{figure*}[t!]
    \centering
    \includegraphics[width=\textwidth]{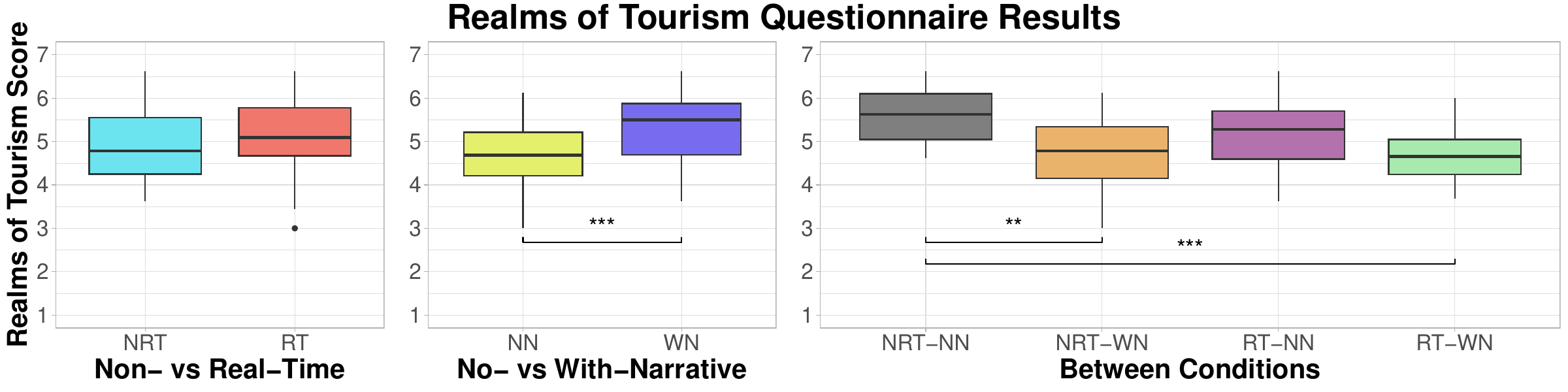}
    \caption{Realms of Tourism score, grouped by: Non-Real-Time vs Real-Time groups (left), No-Narrative vs With-Narrative groups (centre), and by condition (right). * = significant difference.}
    \Description{Three box plots comparing the distributions of the Realms of Tourism scores between groups. The leftmost image compares the Non-Real-Time to the Real-Time group; the distributions are roughly equal and centred at around 4.9. The centre image compares the No-Narrative to the With-Narrative group; the With-Narrative distribution is slightly higher (median = 5.50) than the No-Narrative group (median = 4.69), with a bar drawn between them to indicate that this difference is significant. The rightmost image shows the distributions of the four individual conditions: NRT-NN (median = 4.66), NRT-WN (median = 5.28), RT-NN (median = 4.78), and RT-WN (median = 5.62). There is a bar drawn between NRT-NN and NRT-WN and between NRT-NN and RT-WN to indicate that these distributions are statistically significant.}
    \label{fig:results/rot}
\end{figure*}

The Realms of Tourism were not related to any of our main hypotheses, however we were interested to explore how the presence of narrative or real-time aspects might affect the quality of virtual tours. The mean of each participant's responses to the Realms of Tourism questionnaire were calculated and used as an overall Realms of Tourism (ROT) score which was normally distributed. A two-way ANOVA on the ART-transformed data did not find a significant interaction between narrative and livestreaming for the ROT score ($p = .37$) or for Escapism ($p = .66$), Education ($p = .78$), Esthetics ($p = .87$), Memories ($p = .11$), or Behavioural Intentions ($p = .49$). See \autoref{fig:results/rot} for a graphical depiction of these results or \autoref{tab:ROT} for a full statistical comparison.

Two-way ANOVA found a significant difference in the overall ROT score between the *-WN and *-NN groups ($p < .001$, $\eta^2 = .16$, $PPA = .97$), with the *-WN group ($M = 5.50$, $IQR = 1.19$) scoring significantly higher than the *-NN group ($M = 1.00$, $IQR = 1.00$). A significant difference was also found between these groups for the Escapism ($p < .01$, $\eta^2 = .11$, $PPA = .88$), Education ($p < .0001$, $\eta^2 = .23$, $PPA = 1.00$), Memories ($p < .01$, $\eta^2 = .13$, $PPA = .92$), and Behavioural Intentions subscales ($p < .01$, $\eta^2 = .11$, $PPA = .86$), but not for Esthetics ($p = .41$).

Two-way ANOVA did not find a significant difference between the RT-* and NRT-* groups for overall ROT score ($p = .13$), or for any of the ROT subscales.

%A Wilcoxon signed rank test found a significant difference in ROT score between the No-Narrative and With-Narrative groups ($p < .01$, $r = 0.40$), with the With-Narrative conditions ($M = 5.31$, $SD = 0.78$) scoring significantly higher than the No-Narrative conditions ($M = 4.65$, $SD = 0.81$). Further tests found significant differences in the Education ($p < .001$, $r = 0.51$), Memories ($p = .02$, $r = 0.32$), and Behavioural Intentions ($p = .03$, $r = 0.30$) subscales.% but not in Escapism ($p = .06$) or Esthetics ($p = .40$). No significant difference was found in the ROT score between the Non-Real-Time and Real-Time groups (Wilcox, $p = .29$), or in the realms of Escapism ($p = .34$), Education ($p = .25$), Esthetics ($p = .89$), Memories ($p = .46$), or Behavioural Intentions ($p = .88$).

\section{Discussion}
The aim of this work was to investigate the effect of real-time, narrative-driven virtual tourism experiences on users using a novel fully-immersive telepresence system. This work is relevant in a time when virtual tourism is increasingly considered as a supplement, complement, or even replacement to traditional tourism; an industry that many nations around the world are actively seeking to diversify and move into the digital age. Existing work has called for explorations of meaningful tourism experiences with increased authenticity and application for real-world tourism providers; we hope that our results will help guide these explorations, providing useful ideas for practitioners and future research.

%Finally, the strong correlation we observed between presence and our quality measurements, place attachment and the realms of tourism, suggests that future research should focus on how to increase the sense of presence in the destination environment to create a more engaging experience. 

\subsection{Presence}
Our first hypothesis \textbf{H1} was that the illusion of ``real-time'' travel would increase participants' sense of spatial presence within the virtual environment, as has been observed in previous studies on telepresence~\cite{Yazaki2023, Manabe2020}. However, this was not supported by our results and \textbf{H1} could not be confirmed.

One possible explanation for this is that the deception element was not strong enough to overcome the fact that all participants viewed exactly the same video. While this helped control for possible confounds, it also may have minimised the potential for differences to occur between groups as the only difference between them was purely psychological. This is supported through some of our participants noticing discrepancies between expected and observed behaviour: 
\begin{quote}
       \textit{Participant 48: ``I expected it to be a bit more laggy but there wasn't a lag to his reactions.''}
       
      \textit{Participant 43: ``He wasn't as out of breath as I'd hoped him to be after all that paddling''.} %He was calm, he was nice, he was a good guide. I thought the connection was very stable, I think that really helped that everything was it felt like having a real connection''}
\end{quote}

Despite this scepticism, the deception still seemed to hold, and in contrast with our quantitative results many participants felt as though the real-time aspect was a significant contributor to their sense of presence:
\begin{quote}
    \textit{Participant 55: `Just knowing that was live made it feel a bit more real I think... rather than listening to a prerecorded video, just being a kind of passive viewer rather than an active participant.''}

    \textit{Participant 25: ``I think the important thing for me is like it's real-time, it's not recorded. I think the real time thing makes me feel, like, more real, more connected.''}
\end{quote}
We thus feel confident in our decision to include the deception element and believe that it was necessary for the study's internal validity. However, future research could consider intentionally adding ``technical issues'' to a prerecorded experience, or test a truly real-time system, to determine how this may affect the experience of presence.

%\begin{quote}
        %\textit{``There was very little delay which I was quite impressed by. It made it feel like you could actually have a have a back and forth rather than paging each other, which I thought it was going to be like.''}
    %\end{quote}
 
We did observe significant differences in presence scores between the With-Narrative (*-WN) and No-Narrative (*-NN) groups, with narrative increasing induced presence according to the IPQ score and the General Presence, Spatial Presence, and Involvement subscales. 

%Involvement, the psychological state of focusing one’s energy and attention on the virtual environment~\cite{Schubert2001}, is usually associated with interaction and embodiment; it's possible that speaking with the tour guide in the Real-Time, With-Narrative condition was enough to provide this sense of involvement, however the apparent lack of difference between the Real-Time and Non-Real-Time groups suggests more might be at play.

Our results thus suggest that narrative, regardless of whether it's delivered passively or actively, may facilitate involvement with the experience and increase the sense of presence within it:
\begin{quote}
    \textit{Participant 13: ``I guess it could have been cool if like an audio guide for people like saying what was around or talking about the native flora and fauna.''}
\end{quote}
% We suggest that this is due to the power of narratives to engage users, drawing them into an environment and making them immersed in a story
This is further supported by recent studies on the relationship between media content and presence in VR tourism~\cite{Yu2024, Lee2020.2}, confirming this link can be replicated in a controlled study environment.

Another surprising result is the strong relationship between presence and place attachment and between presence and the realms of tourism. If these are causal relationships then it's not clear in which direction they lie: were participants develop more place attachment to the destination because they felt more spatially present within it? Or did this spatial presence arise because of their attachment to the environment, helping them to forget the real world? Regardless, it seems clear that presence is an important factor in ensuring a satisfying experience for users, and future studies should investigate this link further to determine how best to facilitate a greater connection to the virtual destination.

\subsection{Place Attachment}  
Our second hypothesis \textbf{H2} was that the inclusion of a narrative would significantly increase place attachment to the tourist destination which was supported by our results.

\begin{table}[t!]
    \centering
    \begin{tabular}{ccccc} 
        \toprule
         & \multicolumn{2}{c}{Change in} & \multicolumn{2}{c}{Change in} \\
         & \multicolumn{2}{c}{Pleasure-Displeasure} & \multicolumn{2}{c}{Arousal-Sleepiness} \\
         \cmidrule(lr){2-3}\cmidrule(lr){4-5}
         & $M \pm IQR$& $p$& $M \pm IQR$& $p$\\
         \midrule
         NRT-* & $1.00 \pm 2.00$&                       & \cellcolor{sigcell}$-1.00 \pm 3.50$& \cellcolor{sigcell}                      \\
         RT-*  & $1.00 \pm 2.00$& \multirow{-2}{*}{.26}& \cellcolor{sigcell}$1.00 \pm 3.25$& \multirow{-2}{*}{\cellcolor{sigcell}.0003}\\
         \midrule
         *-NN & $1.00 \pm 1.50$&                       & $-1.00 \pm 3.00$& \\
         *-WN & $1.00 \pm 2.00$& \multirow{-2}{*}{.19}& $0.50 \pm 4.00$& \multirow{-2}{*}{.07}\\
         \bottomrule
    \end{tabular}
    \caption{The Median $M$, Inter-Quartile Range $IQR$, and two-way ANOVA results $p$ for participants' change in Pleasure-Displeasure and Arousal-Sleepiness during the kayaking experience. This change was scored as the difference between a pre-experiment and post-experiment completion of the Affect Grid (see \autoref{fig:affect_grid}, resulting in a possible difference in the range [-8,8]. A negative value indicates a shift towards Displeasure or Sleepiness, while a positive value indicates a shift towards Pleasure or Arousal. Statistically significant differences are highlighted in yellow.}
    \label{tab:Affect}
\end{table}

Despite this, several issues were raised that could affect the validity of our results. First of all, the Place Attachment Scale was not developed for virtual environments, nor has any version of this scale been adapted for VR. As a result, certain items in the questionnaire may not be relevant for measuring PA within virtual environments:
\begin{quote}
    \textit{Participant 45: ``I couldn't exactly answer the [Place Attachment Questionnaire] just because I've only experienced [the place] virtually. But if I did get the experience to go there in person, I would be able to answer more directly about the place I think.''}
\end{quote}

Some participants also noted that the experience simply wasn't long enough for place attachment to form:
\begin{quote}
    \textit{Participant 42: ``I just didn't feel like it was quite long enough to feel that connected... I was, like, just starting to get a sense of the place.''}
    
    \textit{Participant 40: ``I think the experience was a bit too short to get a full grasp of everything.''}
\end{quote}
Fifteen minutes could be considered a short amount of time to spend in a tourist destination but it can be considered long when compared to existing studies on virtual experiences in general.

%Experiences specifically targeting place attachment should thus consider longer run-times, or future research should focus their attention on the aspects of place attachment that can realistically be achieved with such limited exposure to the destination.

Finally, as we opted for semi-random assignment to groups, individual personality differences and/or tourism preferences were not taken into account. For example, some participants found the No-Narrative conditions boring after a few minutes and stopped paying attention:
\begin{quote}
    \textit{Participant 27: ``After the first five minutes it got a bit boring... I'd already looked around in all directions and been like, `oh that's very cool' and then I was sort of just, you know, looking at the same things.''}
\end{quote}
We anticipated that introverted participants may not have enjoyed the RT-WN condition as it required an extended conversation with a stranger, however this did not seem to be the case:
\begin{quote}
    \textit{Participant 25: ``I'm an introvert person, so if I'm truly sitting in a boat, I may just smile to him, I won't, like, talk with him too much. But... he's so kind, and he's trying to explain something to me, and I don't want him to feel, like, sad, so I just try to give him some, like, yeah, feedback.''}

    \textit{Participant 16: ``I'm a quiet person in general. And I was worried, like, oh, I'm going to have to talk to someone.  But it felt really natural.''}
\end{quote}

This is important to note as previous literature has shown that the satisfaction of one's experience in a place is linked to place attachment~\cite{Ramkissoon2015}. If user preference was indeed a factor, it would be important to consider during the design process and should be accommodated, for example by allowing users to toggle the Narrative on or off.   %Importnat to add?: While we attempted to measure the quality of the experience, we did not measure each individual's satisfaction with the experience.

Our results also revealed a significant difference in place attachment between the Non-Real-Time and Real-Time groups. While the reasoning for this isn't clear, we suspect that viewing the destination in real time may increase the sense of having actually visited it, resulting to a greater attachment to the place. This suggests that live-streaming an experience may benefit its quality and reception, however further studies are required to confirm this.

%Nevertheless, we observed a significant difference for one Place Attachment subscale, place identity, between the With-Narrative (*-WN) and No-Narrative (*-NN) groups. As a result, we suspect that narrative may still play a key role in facilitating overall place attachment in VRTEs, however, further research is required to confirm this. We recommend future studies explore specific narrative characteristics including length, narrative type, and emotional framing. 

%%The other sub-dimensions did not reveal significant differences between groups. We were unsurprised by this result for the place dependence sub-dimension, which captures to the utility of a place in one's life. This is likely due to the fact that users did not select the location or activity, increasing the chance that they were irrelevant to the user's life and goals.
%%On the other hand, we were surprised by the results for the final sub-dimension, nature-bonding; which reported no significant differences between groups nor conditions. This was unexpected because our experience and narrative heavily emphasised aspects of the natural environment, including wildlife, plant life, and the ecology of the area. This suggests that either the measure is unable to capture virtually-facilitated nature-bonding, or, that it is not possible to facilitate nature-bonding using virtual reality. Future work should investigate this further. 

\begin{table*}[t!]
    \resizebox{\textwidth}{!}{
    \centering
    \begin{tabular}{ccccccccccccc} 
        \toprule
          & \multicolumn{2}{c}{Realms of Tourism} & \multicolumn{2}{c}{\multirow{2}{*}{Escapism}} & \multicolumn{2}{c}{\multirow{2}{*}{Education}} & \multicolumn{2}{c}{\multirow{2}{*}{Esthetics}} & \multicolumn{2}{c}{\multirow{2}{*}{Memories}} & \multicolumn{2}{c}{Behavioural} \\
          & \multicolumn{2}{c}{Overall} & & & & & & & & & \multicolumn{2}{c}{Intentions} \\
         \cmidrule(lr){2-3}\cmidrule(lr){4-5}\cmidrule(lr){6-7}\cmidrule(lr){8-9}\cmidrule(lr){10-11}\cmidrule(lr){12-13}
         & $M \pm IQR$& $p$& $M \pm IQR$& $p$& $M \pm IQR$& $p$& $M \pm IQR$& $p$& $M \pm IQR$& $p$& $M \pm IQR$& $p$\\
         \midrule
         NRT-* & $4.78 \pm 1.30$& & $4.33 \pm 2.42$& & $5.33 \pm 1.33$& & $4.67 \pm 1.33$& & $4.83 \pm 1.08$& & $5.00 \pm 1.50$& \\
         RT-*  & $5.09 \pm 1.10$& \multirow{-2}{*}{.13}& $5.00 \pm 2.33$& \multirow{-2}{*}{.34} & $5.67 \pm 1.08$& \multirow{-2}{*}{.26} & $4.67 \pm 1.42$& \multirow{-2}{*}{.42}& $5.17 \pm 1.08$& \multirow{-2}{*}{.19}& $5.50 \pm 2.00$& \multirow{-2}{*}{.45}\\
         \midrule
         *-NN & \cellcolor{sigcell}$4.69 \pm 1.00$& \cellcolor{sigcell} & \cellcolor{sigcell}$4.00 \pm 2.33$& \cellcolor{sigcell}& \cellcolor{sigcell}$5.00 \pm 1.33$& \cellcolor{sigcell} & $4.67 \pm 1.33$& & \cellcolor{sigcell}$4.67 \pm 0.83$& \cellcolor{sigcell} & \cellcolor{sigcell}$4.88 \pm 1.40$& \cellcolor{sigcell} \\
         *-WN & \cellcolor{sigcell}$5.50 \pm 1.19$& \multirow{-2}{*}{\cellcolor{sigcell}.0003}& \cellcolor{sigcell}$5.50 \pm 1.67$& \multirow{-2}{*}{\cellcolor{sigcell}.003}& \cellcolor{sigcell}$6.00 \pm 1.33$& \multirow{-2}{*}{\cellcolor{sigcell}.00001}& $4.67 \pm 1.75$& \multirow{-2}{*}{.41}& \cellcolor{sigcell}$5.33 \pm 1.33$& \multirow{-2}{*}{\cellcolor{sigcell}.001}& \cellcolor{sigcell}$5.88 \pm 1.31$& \multirow{-2}{*}{\cellcolor{sigcell}.004}\\
         \bottomrule
    \end{tabular}
    }
    \caption{Median $M$, Inter-Quartile Range $IQR$, and two-way ANOVA results $p$ from the Realms of Tourism questionnaire and its subscales per group. Results are in the range [1,7] with higher results indicating a richer tourism experience. Statistically significant differences are highlighted in yellow.}
    \label{tab:ROT}
\end{table*}

\subsection{The Relationship between Affect and Memory}
Our third hypothesis \textbf{H3} was that participants would be subject to ``rosy retrospection'', or the tendency to remember events more or less fondly based on their change in affect~\cite{Mitchell1997}). This is inspired by previous literature which has demonstrated the brain can make mistakes when encoding memories, even those formed in virtual environments~\cite{Bonnail2024}.

We can accept \textbf{H3} but only for the pleasure-displeasure element of affect, with participants who experienced a decrease in their overall pleasure being significantly more likely to choose a negatively-altered image as a true reflection of their experience, but not to choose a positively-altered one when pleasure is increased. This may be due to a ceiling effect; our results showed that participants were much more likely to choose a positively-altered image regardless of their change in affect, limiting the potential for changes in Pleasure-Displeasure or Arousal-Sleepiness to affect which image was chosen. Future studies should consider how to avoid this bias, possibly by evaluating the positivity of a memory on a continuous scale rather than as a multiclass problem.

Arousal-Sleepiness did not appear to play a role in which image was chosen, supporting prior evidence that arousal and pleasure are distinct~\cite{Loureiro2014}. This suggests that designers wishing for their experience to be remembered fondly should focus on optimising for pleasure, possibly by removing potentially unpleasant aspects of the physical experience. For example, one participant who had physically visited Ōkārito expressed having fonder memories of the virtual tour than their real one:
\begin{quote}
    \textit{Participant 53: ``It was... more peaceful, I guess, quieter... a lot of mosquitoes so it wasn't totally pleasant [when physically] there. This was in some ways nicer, you know''.}
\end{quote}
The role of experience on memory seems like a promising avenue for future research, particularly in VR where user perceptions can be easily manipulated.

\subsection{Arousal}
Unlike the pleasure dimension of affect, arousal did not influence participant's memory of the experience. However, the average change in arousal was significantly different between the Real-Time and Non-Real-Time groups, with those watching a pre-recorded experience becoming more sleepy or relaxed (see \autoref{tab:Affect}).
%Real-time, narrative = higher arousal. 

This relationship reveals that the delivery of the experience, either live or pre-recorded, should be considered depending on its intended effect. For example, some participants enjoyed the peace that a lack of narrative brought:
\begin{quote}
    \textit{Participant 44: ``Once you really get into it, it's quite pleasant. I think I nearly fell asleep because it was so relaxing.''}

    \textit{Participant 53: ``It's very nice to sit here and, you know, be lazy... the water lapping at the sides of the boat. It kind of lulls you almost to sleep a little bit.''}
\end{quote}
Quiet, solo experiences could be used for meditative reflection or relaxation, whereas real-time, guided tours could be used when more engagement is desired:
\begin{quote}
    \textit{Participant 55: ``The interactive effect where you can actually talk to this person who is real, in real time, rather than listening to a prerecorded video... it did definitely make it a more memorable experience.''}

    \textit{Participant 49: ``I think that human element made it really special, unique and memorable. I mean he provided me with so much information that I want to know more and I definitely want to visit. It just added another layer to the experience.''}
\end{quote}

\subsection{The Realms of the Tourism Experience}
The Realms of Tourism have been successfully used to measure the quality of real-world touristic experiences~\cite{Oh2007, Loureiro2014, Song2015}, and so despite not contributing to our hypotheses we were interested in exploring which aspects of the tourism experience could be replicated virtually. Our results showed a significant difference in ROT between the With-Narrative and No-Narrative groups, suggesting that narrative can facilitate a more meaningful experience overall.

Escapism was a particularly strong theme brought up repeatedly by our participants, with the experience having a dissociative effect:
\begin{quote}
    \textit{Participant 56: ``I think the main thing is that I wasn't keeping track of time, right?  Like, in day-to-day life, you keep track of the hours... But when I was there in the kayak, it was like, it doesn't matter. I'm just in a kayak, you know, in this little backwater.''}

    \textit{Participant 19: ``I noticed how quickly you can disconnect yourself and kind of put your mind at ease and just forget everything about what's going on in your life... I feel like this could be a quick getaway for people to enjoy a bit of calm and relaxation in their everyday life.''}
\end{quote}

Lastly, we observed significantly higher levels of Behavioral intention amongst Narrative groups compared to Non-narrative groups. This suggests that the guided tour aspect of the experience made people more likely to speak highly of their experience to others, to recommend it to friends, and increased the likelihood that they would visit the location in real life:
\begin{quote}
    \textit{Participant 30: ``Well, I think it's very good for showing what a place is truly like. Almost like a teaser or a trailer kind of thing.  Yeah, like a tester.''}
\end{quote}
This aligns with prior research which suggests VR is an exceptional tourism marketing tool~\cite{Huang2016}

%These findings provide substantial grounds for further research to build off from. It begs the question: Is maximum presence the ultimate goal for Virtual tourism? Or could other experience aspects be more important, such as arousal or pleasantness of emotion? Ultimately we believe it comes down to the goal of the virtual tourism experience: to manipulate behavioral intentions, to facilitate meaningful and lasting memories, or something else. Therefore, we feels these results contribute significantly and meaningfully to the field of virtual tourism, as they provide concrete evidence that arousal can be increased by real-time systems and storytelling. 
   
\begin{figure}
    \centering
    \includegraphics[width=\linewidth]{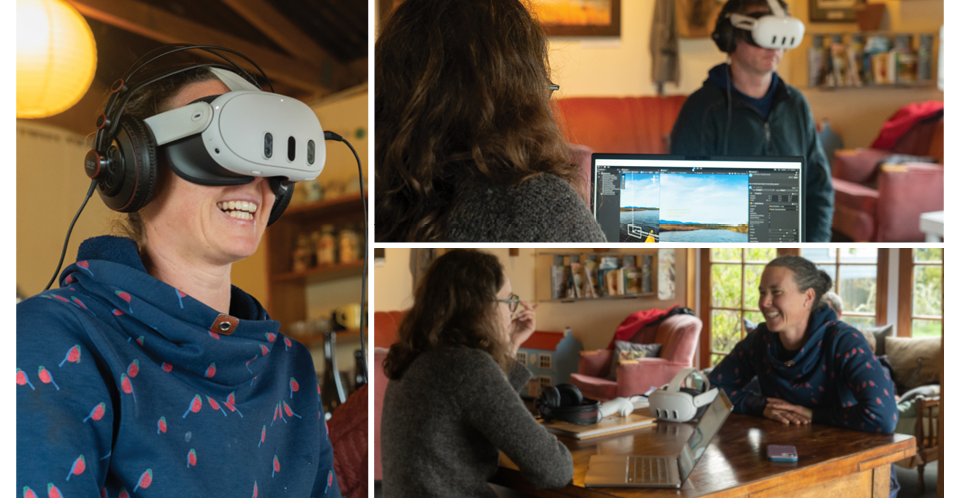}
    \caption{Photos from our demonstration of the final VRTE to our tourism partners. They were shown the full 15-minute experience and gave feedback on the technology and how it could impact their business and the tourism sector.}
    \Description{Three images depicting our demonstration of the final kayaking experience to our tourism partners. The leftmost image is a portrait close-up of a woman smiling while wearing a Meta Quest 3. The other two images are placed to the right of this and stacked on top of each other; the top image is captured over the shoulder of a woman watching the experience on a laptop with a man in the background wearing a Meta Quest 3. The bottom image is also captured over the shoulder of the woman with the laptop, but now she is sitting at a table talking with the woman from the leftmost image while the Meta Quest 3 sits on the table.}
    \label{fig:DomainExperts}
\end{figure}

\subsection{Limitations}
There are several factors that could have impacted our study and its findings. Firstly, our study had 80 participants but used between-subjects/groups with 20 participants per group. While this is rather low, we think our results are robust and internally valid as we limited our scope to rejecting the null hypothesis, and post-hoc power analyses for our main results showed overall reasonable to very good power (0.7-0.96) for our significant findings.

Furthermore, for reasons laid out earlier in our study description, our study used deception for two groups by stating that the experience was live even though it was not. We asked participants during the debrief if they noticed the deception and excluded their results if their interview responses or behaviour indicated they saw through it. However, a chance remains that the deception failed for other participants which may affect the external validity of our results.

Besides these overall limiting factors, a few issues were raised during the post-experiment interviews related to the limitations of current technology. For example, several participants expressed that the lack of sensory elements such as smell and touch may have adversely affected their sense of presence:
\begin{quote}
    \textit{Participant 40: ``I feel like wind and temperature play a huge part in being there... a mild fan would help a lot.''}

%\bn egin{quote}
   %\textit{``I think I would have preferred the real life feeling, like the wind maybe or like the sunlight with heat of the sun on me. It would be really nice.''}
%\end{quote}

    \textit{Participant 55: ``It's not the same as if you went out there and had a trip, you're still lacking some sensory experiences. Part of being in nature is the smells and the feeling of the environment, whether it's windy or there's the sun, you feel the heat. I think those play a big part''}
\end{quote}
%\begin{quote}
    %\textit{``It would be good if you could experience feeling the breeze, feeling the sunlight. That would be really good experience.''}
%\end{quote}

Another point that was raised was that the camera felt unnaturally ``high up'', with several participants expressing that their perspective felt unnatural:
\begin{quote}
    \textit{Participant 36: ``I think that the camera was a bit high, I felt like I was floating.''}

    \textit{Participant 60: ``It would be nice if you were positioned inside the kayak. It was quite unnatural for you to be on top of the kayak, it just felt off.''}
\end{quote}
This was due to the 360$^\circ$ camera being mounted slightly above the kayaker's eye level, which suggests that even small deviations from expected behaviour can reduce the plausibility of the experience.

The blur we applied to the kayaker's face in the 360$^\circ$ video also may have limited presence and seemed to hinder the social connection between participants and the tour guide:
\begin{quote}
    \textit{Participant 26: ``You feel a bit disconnected just for the fact that you're not looking at them, their face is blurred, and it almost feels more like a video chat.''}

    \textit{Participant 49: ``As soon as I was getting immersed in the experience I was turning back and his face was blurred; I was like, oh okay so it's not real.''}
\end{quote}

Despite our attempts to improve the resolution based on feedback from our tourism partners, several participants reported that the 360$^\circ$ camera's low resolution made it difficult to make out details in the environment such as bird life:
\begin{quote}
    \textit{Participant 24: ``I think the resolution could have been better... I could see a splash in the distance but I couldn't see the bird. If it was better it would have made my experience better... as if I was really there.''}

    \textit{Participant 14: ``The quality of the virtual environment still needs to be improved. I can clearly distinguish it from the real world. That's why sometimes I'm actually focused on what is happening outside.''}
\end{quote}
Though this can be solved for prerecorded video with advancements in camera technology, the introduction of video compression for streaming the video live would likely exacerbate this issue further. This is of particular concern in remote areas (such as the lagoon) where network coverage is limited, supporting our decision to limit such factors by prerecording the experience. Future systems that desire a true real-time experience should keep this limitation in mind or could consider a hybrid approach with both prerecorded and live elements.

\subsection{Feedback from Domain Experts}
After conducting our experiment, the research team visited our partners once again to review the results and discuss their future implications for this technology, their business, and the tourism sector in general. We also showed them the final version of the VRTE, including the narrative voice-over, to collect their feedback on the technology and story (see \autoref{fig:DomainExperts}).

Our partners were impressed with the virtual kayaking experience and saw its potential to offer a convenient and accessible means to explore the lagoon:
\begin{quote}
    \textit{``Currently for us to offer an experience to people, really it's a three-hour commitment for them. They have to arrive. They have to get dressed up. They have to go down to the water, do paddle practice with us, go on the water for two hours, come back, tidy up. You could condense that experience down to, for argument's sake, 45 minutes online, where we've stitched together four 10-minute segments. Their experience is condensed, not reliant on them physically having to travel here. And you could present that either as prerecorded or as live. And I'd suggest there'd be tremendous value in that. It would be different to what we offer. And what we offer would still have value; perhaps even deeper value in terms of connection, physical connection to place and to person. But the alternative, the condensed, easier-to-access version could be made really attractive, I think. And could be done really well. I like the idea of that, really, I do.''} 
\end{quote}

They seemed impressed by the visual quality of the 360$^\circ$ video but were skeptical that it could ever match the authentic experience:
\begin{quote}
    \textit{``If you had never been there before, it would seem fantastic in VR. You've got your mountains in the background, you're right up in the forest. I don't know how you could capture it better than what it is. You feel like you're there. It was kind of like, Whoa, never expected it to actually feel this big.''}
    
    \textit{``Zoom tours or Facetime don't reflect the experience at all. They're way too shallow in terms of the impression that the physical place gives, whereas [VR] gets you much, much, much closer to that with the technology. The technology's impressive.''}
\end{quote}

The experience also reinforced their previous belief that virtual technologies could connect more people to Ōkārito than physical travel alone, providing a promising opportunity for small tourism operators to increase their visibility in an accessible and sustainable way: 
\begin{quote}
    \textit{``That's where I'd suggest the opportunity lies for tourism in New Zealand. For someone in Japan to be able to put on that headset and to hear that commentary in Japanese, which the technology should be there and available to do, and to have them feel like they are there. Yeah, that's really where we're overcoming the barriers in tourism that New Zealand faces. It's where the advantage potentially lies, I think.''}
\end{quote}

The condition in which participants were able to speak with the paid actor in real time was of particular interest and identified as a more feasible option than entirely live tours:
\begin{quote}
    \textit{``I'd love to see it more as a like on-demand tour, you know, like [customers watch] the [prerecorded] tour, but you talk to someone in real-time or something like that. You could record a few different days, like, say, one on a different day where it's lower tide and stiller, one when you're more in the main lagoon up into the river channel. You could have a few different [versions] and you could have [the tour guide] live talking''}
\end{quote}

\subsection{Ethical Considerations}
Although we did not aim to study the ethical dimensions of VRT, we collected notable observations from both users and our tourism operators. Most notably, 
despite their general enthusiasm for the experience, our partners also had several concerns about the future of immersive technologies and how they might erode their ability to protect their business and the environment:
\begin{quote}
    \textit{``Once this product was published, what would be the barriers to an AI completely replicating this with none of the benefits flowing to the place? It's currently the [physical] place that protects the uniqueness of the experience. But everything we've talked about, the information we've shared, once a large language model picks that up, it will refine it and refine it and refine it. And it's no longer ours. And we no longer have any ownership or control of it. So it's a significant threat as well in that regard.''}
    
    \textit{``It would be so easy to create another place that actually had higher, clearer mountains in the background, more birds flying around. Always sunny. A better-looking guide behind you. All of those things. A blue kayak instead of yellow. It seems there'd be nothing to stop the experience from being modified and improved, so to speak.''}
\end{quote}

This raises interesting ethical question regarding ownership of virtual replicas (digital twins) of cultural heritage or touristic environments \cite{thompson2017legal} that different from traditional questions of ownership in VR that focus mainly on body and avatars \cite{lin2022digital}. Specifically, how to protect physical and digital property and locations once it's accessible or potentially modifiable to anyone. Or, as our partners succinctly said,
\begin{quote}
    \textit{``Technology is hard to constrain.''}
\end{quote}

Another potential issue arose from the fact that, consistent with previous literature, our participants and tourism operator partners alike praised the technology for its potential to create more sustainable travel experiences, which can be offered more accessibly. Nearly all participants expressed a desire to share this technology with friends or family and a desire to repeat a similar experience. Unfortunately, this excitement also manifested in a pointed desire to travel to the "real" location in-person, which could be directly at odds with the risks of over-tourism. The experience, which we intended to be an on-site experience, seems to unintentionally become a marketing tool, possibly driving more tourism to the location. This outcome may or may not be desirable given the circumstances, but is an ethical consideration for future academics and practitioners nonetheless. 

Lastly, the impact of the experience psychologically has proven to be significant. We have demonstrated that a brief experience, as short as 15-minutes, can profoundly change a visitor's affect and memory. Future studies should further explore the impact of VRTEs on affect and memory in a diverse range of VRTEs which aligns with similar ethical concerns for VR environments raised in the literature \cite{slater2020ethics}.

\section{Conclusion and Future Work}
In this work we explored how real-time and narrative elements within virtual tourism affects the perception of the overall experience. To our knowledge, this is the first systematic analysis of these factors in a tourism context that utilises the expertise of real tourism practitioners. Our findings suggest that narrative is a key element to creating meaningful, memorable VR tourism experiences, resulting in higher presence within and attachment to the destination while increasing the quality of the experience. Live-streaming also appears to play a role, further increasing place attachment to the destination while allowing tourism operators to offer a more flexible and tailor-made experience. 

\balance

Feedback from participants and tourism providers indicates that virtual tourism, as explored in our study, is currently best suited to complement physical travel despite our best efforts to create a meaningful virtual tour that could stand as its own experience. Future work could explore if longer experiences, as suggested by participants, could overcome this impression. We also saw some evidence that personal traits, such as introversion and tourism preferences, are a possible confound for some conditions and might warrant further research for confirmation. Similarly, there are indicators that technological improvements could further improve the experience; at least for the explored context, the ability to clearly distinguish birds seems to be important for the overall experience for both the guide as well as the visitors.

Besides measurements traditionally used in HCI, we applied measurements from the tourism domain, and in doing so identified issues when applying the Abbreviated Place Attachment Scale to virtual environments. This raises the general question of how suitable current measurements from tourism are for their application in VR; we see here potential for future work to validate these measurements and develop a meaningful metric for virtual tourism experiences.

Finally, the digitisation of real environments raises ethical questions regarding ownership of virtual landscapes and environments, and the ability to modify them raises concerns about distorted environment perceptions; while we have seen these concerns raised for other aspects of VR \cite{lin2022digital,slater2020ethics}, to our knowledge this is the first time they have been identified in a tourism context.

%A major take-away from this project is that attempting to recreate realistic, meaningful, live tourism experiences in a style that could replace traditional in-person tourism is likely impossible at this given time. 

As long as there are no social, environmental, or cultural boundaries to travel, people currently might not consider virtual tourism as a viable alternative and still see it as a teaser for the ``real'' experience. We believe our work has shown that emerging immersive technologies have the potential to change this perception, but that it will require a new way of thinking than what is typically applied to tourism. Therefore, we believe future VRT practitioners and researchers should be guided by a different principle: What experiences can one offer in VRTEs that are irreplicable, how can we mitigate the downsides of traditional tourism, and most importantly: what factors are important for creating a high-quality, meaningful experience?

%Our results generally support the notion that the inclusion of narrative in VRTEs can significantly increase presence and place identity, but not overall place attachment. Similarly, narratives significantly increased the quality of the experience and participants' affect within it, leading to an unrealistically positive recollection of the experience. These results imply that narratives are an essential element to creating meaningful VRTEs; though future research should replicate these results using additional contexts/narratives.%, suggesting that narratives may create more meaningful, memorable VRTEs. 

%On the other hand, our results did not demonstrate a significant effect of the ``live'' experience on any of our measurements (presence, place attachment, or overall experience scores), except for change in arousal. This may suggest that a prerecorded experience is as meaningful as a live one, however this requires further validation with a fully/truly real-time system.

\section*{Author Contributions}
\textbf{Lillian Maria Eagan:} Writing - Original Draft, Methodology, Investigation; \textbf{Jacob Young:} Writing - Review \& Editing, Methodology, Software, Supervision, Formal Analysis; \textbf{Jesse Bering:} Writing - Review \& Editing, Supervision, Methodology; \textbf{Tobias Langlotz:} Writing - Review \& Editing, Conceptualisation, Methodology, Funding Acquisition, Project Administration, Supervision.

\begin{acks}
This research was supported by an Endeavour Fund from New Zealand's Ministry of Business, Innovation \& Employment (Contract UOOX2308). We would like to thank Barry and Gemma Hughes for their support and feedback throughout this project, and Rhys Latton for the time and effort he put into acting as the tour guide for our experiments.

\end{acks}

\bibliographystyle{ACM-Reference-Format}
\bibliography{references}

\end{document}